\documentclass[aps,prc,groupedaddress,showpacs]{revtex4}
\usepackage{color,soul}

\usepackage{epsfig}
\usepackage{amssymb}
\usepackage{amsmath}

\newcommand{\del}{\partial}
\newcommand{\beq}{\begin{equation*}}
\newcommand{\eeq}{\end{equation*}}
\newcommand{\be}{\begin{equation}}
\newcommand{\ee}{\end{equation}}
\newcommand{\beqa}{\begin{eqnarray}}
\newcommand{\eeqa}{\end{eqnarray}}
\newcommand{\bea}{\begin{eqnarray}}
\newcommand{\eea}{\end{eqnarray}}
\newcommand{\bra}{\langle}
\newcommand{\ket}{\rangle}

\newcommand{\s}{\sigma}
\newcommand{\w}{\omega}
\newcommand{\reci}[1]{\frac{1}{#1}}

\begin{document}

%\preprint{}

\title{Phase Transition from QMC Hyperonic Matter to Deconfined Quark Matter}


%%\author{me}
%\thanks{}
%%\affiliation{here}

\author{J.~D. Carroll} 
\email{jcarroll@physics.adelaide.edu.au}
\author{D.~B. Leinweber}
\author{A.~G. Williams}

\affiliation{Centre for the Subatomic Structure of Matter (CSSM),
Department of Physics, University of Adelaide, SA 5005, Australia}

\author{A.~W. Thomas}
\affiliation{Thomas Jefferson National Accelerator Facility, 
12000 Jefferson Ave., Newport News, VA 23606, USA}
\affiliation{College of William and Mary, Williamsburg, VA 23187, USA.}
\affiliation{Centre for the Subatomic Structure of Matter (CSSM),
Department of Physics, University of Adelaide, SA 5005, Australia}


\date{\today}

\begin{abstract}

We investigate the possibility and consequences of phase transitions
from an equation of state (EoS) describing nucleons and hyperons
interacting via mean-fields of $\s$, $\w$, and $\rho$ mesons in the
recently improved Quark-Meson Coupling (QMC) model to an EoS
describing a Fermi gas of quarks in an MIT bag.  The transition to a
mixed phase of baryons and deconfined quarks, and subsequently to a
pure deconfined quark phase is described using the method of
Glendenning.  The overall EoS for the three phases is calculated for
various scenarios and these are used to calculate stellar solutions
using the Tolman-Oppenheimer-Volkoff equations. The results are
compared to recent experimental data and the validity of each case is
discussed with consequences for determining the species content of the
interior of neutron stars.
\end{abstract}

%\keywords{}

\maketitle

% main text

\section{Introduction}
\label{sec:intro}

The use of hadronic models to describe high density matter enables us
to investigate both the microscopic world of atomic nuclei and the
macroscopic world of compact stellar objects, encompassing an enormous
range of scales. The results of these investigations provide deep
fundamental insight into the way that the world around us is
constructed.\par

Experimental data from both extremes of scale aid in constraining such
models, from the saturation properties of nuclear matter to the
observed properties of neutron
stars~\cite{Podsiadlowski:2005ig,Grigorian:2006pu,Klahn:2006iw}.  The
literature~\cite{Lattimer:2000nx,Heiselberg:1999mq,Weber:2004kj,SchaffnerBielich:2004ch,Weber:1989hr,Chin:1974sa}
provides a plethora of models for the EoS of hadronic matter, at least
some of which have been successfully applied to calculate the
properties of finite nuclei. There are also important constraints from
data involving heavy-ion
collisions~\cite{Danielewicz:2002pu,Worley:2008cb}.  Many of these EoS
have also been applied to neutron star features as well. However, the
amount of data available for neutron stars (or compact stellar
objects) is very limited, with only a single result containing both a
mass and radius result simultaneously~\cite{Ozel:2006bv} and even that
has been recently disputed~\cite{Alford:2006vz}.\par

With such a lack of constraining data, our focus shifts to finding
models which better reflect the physics that is expected to be
important under the conditions which we are investigating. A prime
example of this is that at the densities which we consider to be
interesting for this investigation (1--10 times nuclear density,
$\rho_0 = 0.16~{\rm fm}^{-3}$) it is possible that either hyperonic
matter (in which strangeness carrying baryons become energetically
favourable as the Fermi sea fills), quark matter (in which it becomes
energetically favourable that the quarks inside the baryons become
deconfined) or a mixed phase of these is present, rather than the more
traditional treatment of nucleons alone.\par

We construct a model of high density matter which is globally charge
neutral, color neutral, rotationally symmetric, and in a state that is
energetically favourable. For this purpose we consider hadronic matter
modelled by the Quark-Meson Coupling (QMC)
model~\cite{Guichon:2004xg,Guichon:2006er}, which was recently
improved through the self-consistent inclusion of color hyperfine
interactions~\cite{RikovskaStone:2006ta}.  While this improvement had
no significant effect on the binding of nucleons, it led to impressive
results for finite hypernuclei~\cite{Guichon:2007ru}.  We follow the
method of Glendenning~\cite{Glendenning:2001pe} to produce a mixed
phase of hyperonic matter and deconfined quark matter under total
mechanical stability, then a pure deconfined quark matter phase with
relativistic non-interacting quarks.\par

We begin with a brief presentation of Relativistic Mean-Field Theory
in Section~\ref{sec:QHD} to establish a foundation with which to
discuss the general formalism for the QMC model, including new
additions, in Section~\ref{sec:QMC}. Deconfined quark matter is
discussed in Section~\ref{sec:quarkmatter}. This is followed by
Section~\ref{sec:PT} providing a summary of the requirements for and
method to construct a phase transition from a hadronic phase to a
mixed phase and from a mixed phase to a quark phase.  Stellar
solutions are calculated in Section~\ref{sec:TOV} and a summary of our
results is presented in Section~\ref{sec:results} with conclusions in
Section~\ref{sec:conc}.\par

\section{Relativistic Mean-Field Theory} \label{sec:QHD}

We introduce the mean-field description of nuclear matter using the
classic example of Quantum Hadrodynamics
(QHD)~\cite{Chin:1974sa,Serot:1984ey,Furnstahl:2000in}.  Although the
Quark-Meson Coupling model (QMC) has a fundamentally different
starting point, namely the self-consistent modification of the
structure of a hadron immersed in the nuclear
medium~\cite{Guichon:1987jp,Saito:1996yb,Saito:2005rv}, in practice
the equations for nuclear matter involve only a few key
differences. We summarize those in the next section. The original
formulation of QHD included only nucleons interacting with
scalar-isoscalar, $\s$, and vector-isoscalar, $\w$, mesons. This was
later expanded to include the vector-isovector, $\rho$, and
subsequently the entire octet of baryons, \mbox{$B \in \{
  p,n,\Lambda,\Sigma^+,\Sigma^0,\Sigma^-,\Xi^0,\Xi^- \}$}, with global
charge neutrality upheld via leptons, $\ell \in \{e^-,\mu^-\}$.\par

The Lagrangian density for QHD is
\bea 
\nonumber
&{\cal L}\ = & \ \sum_k \bar{\psi}_k \left[\gamma_{\mu}(i\del^{\mu}-g_{\w k}\omega^{\mu}-
g_{\rho}\vec{\tau}_{(k)}\cdot\vec{\rho}^{\, \mu})-(M_k-{g_{\s k}} \s)\right]\psi_k \\
\label{eq:QHDlag}
&& + \frac{1}{2}(\del_{\mu}\sigma\del^{\mu}\sigma-m^{2}_{\s}\sigma^{2})
- \frac{1}{4}F_{\mu\nu}F^{\mu\nu} - \frac{1}{4}R^a_{\mu\nu}R_a^{\mu\nu} \\
\nonumber
&& + \frac{1}{2}m^{2}_{\omega}\omega_{\mu}\omega^{\mu}+\frac{1}{2}m^{2}_{\rho}\rho^a_{\mu}\rho_a^{\mu}
+ \bar{\psi}_\ell\left[\gamma_\mu i \del^\mu - m_\ell \right] \psi_\ell  +\delta\mathcal{L},
\eea
where the index $k\in\{N,\Lambda,\Sigma,\Xi\}$ represents each isospin
group of the baryon states, and $\psi_k$ corresponds to the Dirac
spinors for these
\be
\psi_N = \begin{pmatrix}\psi_p\\\psi_n\end{pmatrix}, \quad
\psi_\Lambda = \begin{pmatrix}\psi_\Lambda\end{pmatrix}, \quad
\psi_\Sigma = \begin{pmatrix}\psi_{\Sigma^+}\\\psi_{\Sigma^0}\\\psi_{\Sigma^-}\end{pmatrix}, \quad
\psi_\Xi = \begin{pmatrix}\psi_{\Xi^0}\\\psi_{\Xi^-}\end{pmatrix}. \quad
\ee
The vector field tensors are
\be \label{eq:vecfieldtensors}
F^{\mu\nu} = \del^\mu\w^\nu - \del^\nu\w^\mu, \quad 
R_a^{\mu\nu} = \del^\mu\rho_a^\nu - \del^\nu\rho_a^\mu 
- g_{\rho}\epsilon^{abc} \rho_b^\mu\rho_c^\nu,
\ee
The third components of the isospin matrices are
\be
\tau_{(N)3} = \tau_{(\Xi)3} = \reci{2} \left[ \begin{matrix}\ 1\ & 0\ \\\ 0\ & -1\ \end{matrix} \right], \quad
\tau_{(\Lambda)3} = \left[ \begin{matrix}\ 0\ \ \end{matrix} \right], \quad
\tau_{(\Sigma)3} = \left[ \begin{matrix}\ 1\ &\ 0\ &\ 0\ \\\ 0\ &\ 0\ &\ 0\ \\\ 0\ &\ 0\ & -1\ \end{matrix} \right], \quad
\ee
$\psi_\ell$ is a spinor for the lepton states, and $\delta{\cal L}$
are renormalisation terms. We do not include pions here as they
provide no contribution to the mean-field, because the ground state of
nuclear matter is parity-even. We have neglected nonlinear meson terms
in this description for comparison purposes, though it has been shown
that the inclusion of nonlinear scalar meson terms produces a
framework consistent with the QMC model without the added hyperfine
interaction~\cite{Muller:1997re}.  The values of the baryon and meson
masses in vacuum are summarized in Table~\ref{tab:masses}.\par

\begin{table}[b]
\centering
\caption{\protect\label{tab:masses}The vacuum (physical) baryon and
  meson masses (in units of MeV) as used
  here~\cite{Yao:2006px}.\vspace{2mm}}
\begin{ruledtabular}
%\begin{tabular}{\hsize}{@{\extracolsep{\fill}}cccccccc}
\begin{tabular}{cccccccc}
%\hline
%\hline
$M_{p}$         & 
$M_n$           &
$M_\Lambda$     & 
$M_{\Sigma^-}$  & 
$M_{\Sigma^0}$  & 
$M_{\Sigma^+}$  & 
$M_{\Xi^-}$     &
$M_{\Xi^0}$     \\
938.27          &
939.57          &
1115.68         &
1197.45         &
1192.64         &
1189.37         &
1321.31         &
1314.83         \\
\hline
\multicolumn{3}{c}{$m_\s$}   &
\multicolumn{2}{c}{$m_\w$}   & 
\multicolumn{3}{c}{$m_\rho$} \\
\multicolumn{3}{c}{$550.0$}  &
\multicolumn{2}{c}{$782.6$}  & 
\multicolumn{3}{c}{$775.8$}  \\
%\hline
%\hline
\end{tabular}
\end{ruledtabular}
\end{table}

Assuming that the baryon density is sufficiently large, we use a
Mean-Field Approximation (MFA) with physical parameters (breaking
charge symmetry) in which the meson fields are replaced by their
classical vacuum expectation values. With this condition, the
renormalisation terms can be neglected.\par

By enforcing rotational symmetry and working in the frame where the
matter as a whole is at rest, we set all of the 3-vector components of
the vector meson fields to zero, leaving only the temporal
components. Furthermore, by enforcing isospin symmetry we remove all
charged meson states.  Consequently, because the mean-fields are
constant, all meson derivative terms vanish, and thus so do the vector
field tensors.  The only non-zero components of the vector meson mean
fields are then the time components, $\bra\w^\mu\ket =
\bra\w\ket\delta^{\mu 0}$ and $\bra\rho^\mu\ket =
\bra\rho\ket\delta^{\mu 0}$.  Similarly, only the third component of
the $\rho$ meson mean field in iso-space is non-zero, corresponding to
the uncharged $\rho$ meson.\par

The couplings of the mesons to the baryons are found via SU(6)
flavor-symmetry~\cite{Rijken:1998yy}. This produces the following
relations for the $\s$ and $\w$ couplings to each isospin group (and
hence each baryon $B$ in that isospin group)
\be
\reci{3}\; g_{\s N} = \reci{2}\; g_{\s \Lambda} = \reci{2}\; g_{\s \Sigma} = g_{\s \Xi},
\quad
\reci{3}\; g_{\w N} = \reci{2}\; g_{\w \Lambda} = \reci{2}\; g_{\w \Sigma} = g_{\w \Xi}.
\ee
Using the formalism as above with isospin expressed explicitly in the
Lagrangian density, the couplings of the $\rho$ meson to the octet
baryons are unified, thus
%and are related to the $\w$ coupling via
%
%\be
%g_{\rho} = \reci{3}\; g_{\w N}
%\ee
%
by specifying $g_{\s N}$, $g_{\w N}$, and $g_{\rho}$ we are therefore
able to determine the couplings to the remaining baryons.\par
 
By evaluating the equations of motion from the Euler-Lagrange equations
\be \label{eq:EL}
\frac{\del\mathcal{L}}{\del\phi_i} 
- \del_{\mu}\frac{\del\mathcal{L}}{\del(\del_{\mu}\phi_i)} = 0,
\ee
we find the mean-field equations for each of the mesons, 
as well as the baryons. The equations for the meson fields are
\bea \label{eq:MFsigma}
&\bra\s\ket &= \sum_B \frac{g_{\s B}}{m^{2}_{\s}}\bra\bar{\psi}_B\psi_B\ket, \\
\label{eq:MFomega}
&\bra\w\ket &= \sum_B \frac{g_{\w B}}{m^{2}_\w}\bra\bar{\psi}_B\gamma^{0}\psi_B\ket 
= \sum_B \frac{g_{\w B}}{m^{2}_{\w}}\bra\psi_B^\dag \psi_B\ket, \\
\label{eq:MFrho}
&\bra\rho\ket &= \sum_k \frac{g_{\rho}}{m^{2}_{\rho}}\bra\bar{\psi}_k\gamma^{0} \tau_{(k)3} \psi_k\ket 
= \sum_k \frac{g_{\rho}}{m^{2}_{\rho}}\bra\psi_k^\dag  \tau_{(k)3} \psi_k\ket 
= \sum_B \frac{g_{\rho}}{m^{2}_{\rho}}\bra\psi_B^\dag I_{3B} \psi_B\ket ,
\eea
where the sum over $B$ corresponds to the sum over the octet baryon
states, and the sum over $k$ corresponds to the sum over isospin
groups. $I_{3B}$ is the third component of isospin of baryon $B$, as
found in the diagonal elements of $\tau_{(k)3}$.  $\bra\omega\ket$,
$\bra\rho\ket$, and $\bra\sigma\ket$ are proportional to the conserved
baryon density, isospin density and scalar density respectively, where
the scalar density is calculated self-consistently.\par

The Euler-Lagrange equations also provide a Dirac equation for the
baryons
\be \label{eq:dirac}
\sum_B \left[i\!\not\!\partial-g_{\w B}\gamma^0\bra\w\ket-
g_{\rho}\gamma^0I_{3B}\bra\rho\ket - M_B + g_{\s B} \bra\s\ket \right]\psi_B = 0.
\ee
At this point, we can define the baryon effective mass as
\be \label{eq:effM}
M_B^* = M_B - g_{\s B} \bra\s\ket,
\ee
and the baryon chemical potential (also known as the Fermi energy, the
energy associated with the Dirac equation) as
\be \label{eq:mu}
\mu_B = \epsilon_{F_B} = \sqrt{k_{F_B}^2 + (M_B^*)^2}+g_{\w B}\bra\w\ket + g_{\rho}I_{3B}\bra\rho\ket.
\ee
The chemical potentials for the leptons are found via
\be \label{eq:ellmu}
\mu_\ell = \sqrt{k_{F_\ell}^2 + m_\ell^2}.
\ee
The energy density, ${\cal E}$, and pressure, $P$, for the EoS can be
obtained using the relations for the energy-momentum tensor (where
$u^\mu$ is the 4-velocity)
\be \label{eq:SET}
\bra T^{\mu \nu}\ket = \left({\cal E}+P\right)u^{\mu}u^{\nu} + P {\rm g}^{\mu\nu},\quad \Rightarrow \quad
P = \reci{3} \bra T^{ii} \ket,\quad {\cal E} = \bra T^{00} \ket,
\ee
since $u^i=0$ and $u_0 u^0=-1$, where ${\rm g}^{\mu\nu}$ here is the
inverse metric tensor having a negative temporal component, and
$T^{\mu\nu}$ is the energy-momentum tensor.  In accordance with
Noether's Theorem, the relation between the energy momentum tensor and
the Lagrangian density is
\be \label{eq:EML}
T^{\mu \nu} = -{\rm g}^{\mu \nu}\mathcal{L}+\del^{\mu}\psi \frac{\del \mathcal{L}}{\del (\del_{\nu}\psi)},
\ee
and we find the Hartree-level energy density and pressure for the
system as a sum of contributions from baryons, $B$; leptons, $\ell$;
and mesons, $m$ to be
\bea \label{eq:E_H}
\nonumber
&{\cal E} &= \sum_{j=B,\ell,m} {\cal E}_j \\
\nonumber
&&= \sum_{i=B,\ell} \frac{\left(2J_i+1\right)}{(2 \pi)^3}
\int \theta (k_{F_i} - |\vec{k}|) \sqrt{k^2+(M_i^*)^2} \; d^3k 
+ \sum_{\alpha = \s,\w,\rho} \reci{2} m_\alpha^2\bra\alpha\ket^2,\\
&& \\
\nonumber
&P &= \sum_{j=B,\ell,m} P_j \\
\nonumber
&& = \sum_{i=B,\ell} \frac{\left(2J_i+1\right)}{3(2 \pi)^3}
\int
\frac{k^2 \; \theta (k_{F_i} - |\vec{k}|)}{\sqrt{k^2+(M_i^*)^2}} \; d^3k
+ \sum_{\alpha = \w,\rho} \reci{2} m_\alpha^2\bra\alpha\ket^2 
- \reci{2} m_\s^2\bra\s\ket^2,\\
\label{eq:P_H}
&&
\eea
where $J_i$ is the spin of particle $i$ ($J_i=\frac{1}{2}\ \forall\ i
\in \{B,\ell\}$) which in this case accounts for the availability of
both up and down spin-states. $\theta(x)$ is the Heaviside Step
Function.  Note that the pressure arising from the vector mesons is
positive, while it is negative for the scalar meson.\par

The total baryon density, $\rho$, can be calculated via
\be \label{eq:rho}
\rho = \sum_{B} \rho_B = \sum_B \frac{\left(2J_B+1\right)}{(2\pi)^3}
\int \theta (k_{F_B} - |\vec{k}|)\; d^3k,% = \frac{2k_{F}^3}{3\pi^3},
\ee
where in symmetric matter, the Fermi momenta are related via
\mbox{$k_F = k_{F_n} = k_{F_p}$}, and the binding energy per baryon,
$E$, is determined via
\be \label{eq:EperA}
E = \left[ \reci{\rho} \left( {\cal E} -\sum_B M_B\rho_B\right) \right].
\ee
The couplings $g_{\s N}$ and $g_{\w N}$ are determined such that
symmetric nuclear matter \mbox{(in which $\rho_p = \rho_n = 0.5\rho$)}
saturates with the appropriate minimum in the binding energy per
baryon of \mbox{$E_0 = -15.86~{\rm MeV}$} at a nuclear density of
\mbox{$\rho_0 = 0.16~{\rm fm}^{-3}$}. The couplings for QHD which
provide a fit to saturated nuclear matter are shown in
Table~\ref{tab:QHDparams}.\par

The coupling $g_{\rho}$ is fixed such that the nucleon symmetry
energy, given by
\be \label{eq:a4}
a_{\rm sym} = \frac{g_{\rho}^2}{12 \pi^2 m_\rho^2} k_{F}^3 
+ \reci{12} \frac{k_{F}^2}{\sqrt{k_{F}^2 + (M_p^*)^2}} 
+ \reci{12} \frac{k_{F}^2}{\sqrt{k_{F}^2 + (M_n^*)^2}} \, , 
\ee
is reproduced at saturation as $(a_{\rm sym})_0 = 32.5~{\rm MeV}$.\par

The chemical potential for any particle, $\mu_i$, can be related to
two independent chemical potentials --- we choose that of the neutron
($\mu_n$) and the electron ($\mu_e$) --- and thus we use a general
relation
\be
\label{eq:chempotrel}
\mu_i = B_i \mu_n - Q_i \mu_e \ ; \quad \ i \in \{ p,n,\Lambda,\Sigma^+,\Sigma^0,\Sigma^-,\Xi^0,\Xi^-,\ell \},
\ee
where $B_i$ and $Q_i$ are the baryon (unitless) and electric (in units
of the proton charge) charges respectively. For example, the proton
has $B_p = +1$ and $Q_p = +1$, so it must satisfy $\mu_p = \mu_n -
\mu_e$ which is familiar as $\beta$-equilibrium.  Since neutrinos are
able to escape the star, we consider $\mu_\nu = 0$.  Leptons have
$B_\ell=0$, and all baryons have $B_B=+1$.\par

The relations between the chemical potentials are therefore derived to
be
\be \label{eq:allmus}
\begin{array}{rl}
\mu_\Lambda = \mu_{\Sigma^0} = \mu_{\Xi^0}\  &= \mu_n, \\
\mu_{\Sigma^-} = \mu_{\Xi^-}\  &= \mu_n + \mu_e, \\
\mu_p = \mu_{\Sigma^+}\  &= \mu_n - \mu_e, \\
\mu_\mu\  &= \mu_e.
\end{array}
\ee
The EoS for QHD can be obtained by finding solutions to
Eqs.~(\ref{eq:MFsigma}--\ref{eq:MFrho}) subject to charge neutrality,
conservation of a chosen total baryon number, and equivalence of
chemical potentials. These conditions can be summarised as
\be 
\label{eq:equilconds}
\left.
\begin{array}{rl}
0 &= \sum_i Q_i \rho_i\\
\rho &= \sum_i B_i \rho_i\\ 
\mu_i &= B_i \mu_n - Q_i \mu_e
\end{array}
\quad \right\} \quad i \in \{ p,n,\Lambda,\Sigma^+,\Sigma^0,\Sigma^-,\Xi^0,\Xi^-,\ell \}.
\ee
With these conditions, we are able to find the EoS for QHD. It should
be noted that, as with many relativistic models for baryonic matter,
once we include more than one species of baryon this model eventually
produces baryons with negative effective masses at sufficiently high
densities ($\rho > 1~{\rm fm}^{-3}$).  This is a direct result of the
linear nature of the effective mass as shown in Eq.~(\ref{eq:effM}).
As the Fermi energy (see Eq.~(\ref{eq:mu})) approaches zero, the cost
associated with producing baryon-anti-baryon pairs is reduced and at
this point the model breaks down.  From a more physical point of view,
as the density rises one would expect that the internal structure of
the baryons should play a role in the dynamics. Indeed, within the QMC
model, the response of the internal structure of the baryons to the
applied mean scalar field ensures that no baryon mass ever becomes
negative. We now describe the essential changes associated with the
QMC model.\par

\begin{table}[!h]
\centering
\caption{\protect\label{tab:QHDparams} Couplings for QHD with the
  octet of baryons, fit to saturation of nuclear matter.\vspace{2mm}}
\begin{ruledtabular}
%\begin{tabular*}{\hsize}{@{\extracolsep{\fill}}ccccc}
\begin{tabular}{ccccc}
%\hline
%\hline
\phantom{space} & $g_{\s N}$ & $g_{\w N}$ & $g_{\rho}$ & \phantom{space} \\
& 10.644 & 13.179 & 6.976 &  \\
%\hline
%\hline
\end{tabular}
\end{ruledtabular}
\end{table}

\section{QMC model} \label{sec:QMC}

Like QHD, QMC is a relativistic quantum field theory formulated in
terms of the exchange of scalar and vector mesons. However, in
contrast with QHD these mesons couple not to structureless baryons but
to clusters of confined quarks. As the density of the medium grows and
the mean scalar and vector fields grow, the structure of the clusters
adjusts self-consistently in response to the mean-field coupling.
While such a model would be extremely complicated to solve in general,
it has been shown by Guichon {\it et al.}~\cite{Guichon:1995ue} that
in finite nuclei one should expect the Born-Oppenheimer approximation
to be good at the 3\% level. Of course, in nuclear matter it is exact
at mean-field level.\par

Within the Born-Oppenheimer approximation, the major effect of
including the structure of the baryon is that the internal quark wave
functions respond in a way that opposes the applied scalar field. To a
very good approximation this physics is described through the ``scalar
polarizability,'' $d$, which in analogy with the electric
polarizability describes the term in the baryon effective mass
quadratic in the applied scalar
field~\cite{Guichon:1987jp,Thomas:2004iw,Ericson:2008tv,Massot:2008pf,Chanfray:2003rs}.
Recent explicit calculations of the equivalent energy functional for
the QMC model have demonstrated the very natural link between the
existence of the scalar polarizability and the many-body forces, or
equivalently the density dependence, associated with successful,
phenomenological forces of the Skyrme
type~\cite{Guichon:2004xg,Guichon:2006er}.  In nuclear matter the
scalar polarizability is the {\it only} effect of the internal
structure in mean-field approximation. On the other hand, in finite
nuclei the variation of the vector field across the hadronic volume
also leads to a spin-orbit term in the nucleon
energy~\cite{Guichon:1995ue}.\par

Once one chooses a quark model for the baryons, and specifies the
quark-level meson couplings, there are no new parameters associated
with introducing any species of baryon into the nuclear matter.  Given
the well known lack of experimental constraints on the forces between
nucleons and hyperons, let alone hyperons and hyperons, which will be
of great practical importance as the nuclear density rises above
(2--3)$\rho_0$, this is a particularly attractive feature of the QMC
approach and it is crucial for our current investigation. Indeed, we
point to the very exciting recent results of the QMC model, modified
to include the effect of the scalar field on the hyperfine
interaction, which led to $\Lambda$ hypernuclei being bound in quite
good agreement with experiment and $\Sigma$ hypernuclei being unbound
because of the modification of the hyperfine
interaction~\cite{Guichon:2007ru} - thus yielding a very natural
explanation of this observed fact. We note the success that this
description has found for finite nuclei as noted
in~\cite{Guichon:2006er}.\par

While we focus on the MIT bag model~\cite{Chodos:1974je} as our
approximation to baryon structure, we note that there has been a
parallel development~\cite{Bentz:2001vc} based upon the covariant,
chiral symmetric NJL model~\cite{Nambu:1961tp}, with quark confinement
modelled using the proper time regularization proposed by the
T\"ubingen group~\cite{Hellstern:1997nv,Ebert:1996vx}.  The latter
model has many advantages for the computation of the medium
modification of form factors and structure functions, with the results
for spin structure functions~\cite{Cloet:2005rt,Cloet:2006bq} offering
a unique opportunity to test the fundamental idea of the QMC model
experimentally.  However, in both models it is the effect of quark
confinement that leads to a positive polarizability and a natural
saturation mechanism.\par

Although the underlying physics of QHD and QMC is rather different, at
the hadronic level the equations to be solved are very similar.  We
therefore focus on the changes which are required.\par

{\bf 1.} \quad Because of the scalar polarizability of the hadrons,
which accounts for the self-consistent response of the internal quark
structure of the baryon to the applied scalar
field~\cite{Guichon:2006er}, the effective masses appearing in QMC are
non-linear in the mean $\sigma$ field.  We write them in the general
form
\be \label{eq:effMQMC}
M_B^* = M_B - w_B^\s \; g_{\s N} \bra\s\ket 
+ \frac{d}{2} \tilde{w}_B^\s \; (g_{\s N}\bra\s\ket)^2
\, , 
\ee
where the weightings, $w_B^\s,\ \tilde{w}_B^\s,$ and the scalar
polarizability of the nucleon, $d$, must be calculated from the
underlying quark model. Note now that only the coupling to the
nucleons, $g_{\s N}$, is required to determine all the effective
masses.\par

The most recent calculation of these effective masses, including the
in-medium dependence of the spin dependent hyperfine
interaction~\cite{Guichon:2007ru}, yields the explicit expressions:
\begin{eqnarray} \label{eq:MstarsinQMC}
M_{N}(\bra\s\ket) & = & M_{N}-g_{\s N}\bra\s\ket\nonumber \\\nonumber
& &+\left[0.0022+0.1055R_{N}^{\rm free}-
0.0178\left(R_{N}^{\rm free}\right)^{2}\right]
\left(g_{\s N}\bra\s\ket\right)^{2},
\label{eq:A18}\\
M_{\Lambda}(\bra\s\ket) & = & M_{\Lambda}-\left[0.6672+0.0462R_{N}^{\rm free}-
0.0021\left(R_{N}^{\rm free}\right)^{2}\right]g_{\s N}\bra\s\ket
\nonumber \\ \nonumber
 &  & +\left[0.0016+0.0686R_{N}^{\rm free}-0.0084\left(R_{N}^{\rm free}\right)^{2}
\right]\left(g_{\s N}\bra\s\ket\right)^{2},
\\ \label{eq:A19usethis}
M_{\Sigma}(\bra\s\ket) & = & M_{\Sigma}-\left[0.6706-0.0638R_{N}^{\rm free}-
0.008\left(R_{N}^{\rm free}\right)^{2}\right]g_{\s N}\bra\s\ket
 \\\nonumber
 &  & +\left[-0.0007+0.0786R_{N}^{\rm free}-0.0181\left(R_{N}^{\rm free}\right)^{2}
\right]\left(g_{\s N}\bra\s\ket\right)^{2},
\label{eq:A21}\\
M_{\Xi}(\bra\s\ket) & = & M_{\Xi}-\left[0.3395+0.02822R_{N}^{\rm free}-
0.0128\left(R_{N}^{\rm free}\right)^{2}\right]g_{\s N}\bra\s\ket
\nonumber \\\nonumber
 &  & +\left[-0.0014+0.0416R_{N}^{\rm free}-0.0061\left(R_{N}^{\rm free}\right)^{2}
\right]\left(g_{\s N}\bra\s\ket \right)^{2}\, . 
\label{eq:A22}
\end{eqnarray}
We take $R_N^{\rm free}=0.8~{\rm fm}$ as the preferred value of the
free nucleon radius, although in practice the numerical results depend
only very weakly on this parameter~\cite{Guichon:2006er}.\par

\begin{figure}[!t]
\centering \includegraphics[angle=90,width=0.9\textwidth]{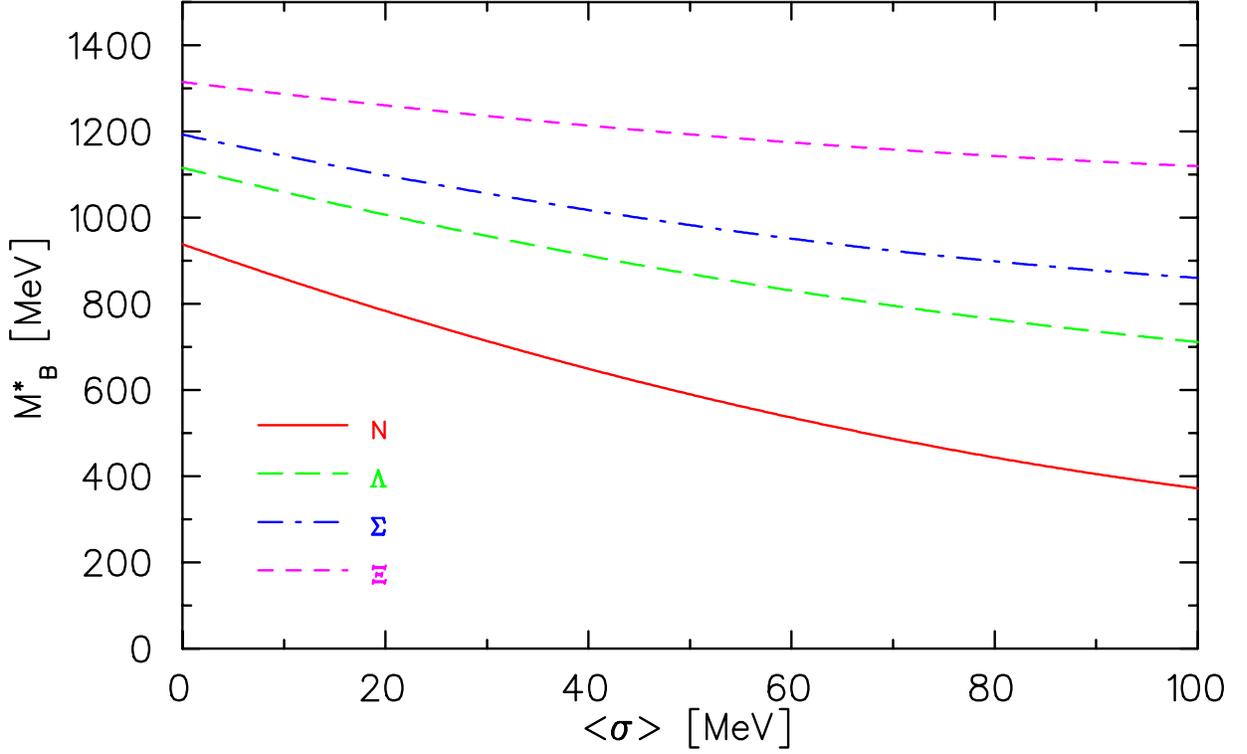}
\caption{(Color online) Baryon effective masses within the QMC model,
  parameterized as a function of the mean scalar field, $\bra\s\ket$.
  The values at $\bra\s\ket=0$ are the vacuum masses as found in
  Table~\ref{tab:masses}.  We show the effective masses only up to
  $\bra\s\ket = 100~{\rm MeV}$ which corresponds to about $2~{\rm
    fm}^{-3}$ (6--8 $\rho_0$), beyond which higher order terms not
  shown in Eq.~(\ref{eq:effMQMC}) become significant.
  \protect\label{fig:effMvsS}}
\end{figure}

Given the parameters in Eq.~(\ref{eq:MstarsinQMC}), all the effective
masses for the baryon octet are entirely determined. They are plotted
as functions of $\bra\s\ket$ in Fig.~\ref{fig:effMvsS} and we see
clearly that they never become negative.  (Note that the range of
$\bra\s\ket$ covered here corresponds to densities up to
(6--8)$\rho_0$).\par

{\bf 2.} \quad Since the mean scalar field, $\bra\s\ket$, is derived
self-consistently by taking the derivative of the energy density with
respect to $\bra\s\ket$, the scalar field equation
\be \label{eq:scalarfield}
\bra\s\ket = \sum_B \frac{g_{\s N}}{m_\s^2} C(\bra\s\ket) 
\frac{\left(2J_B+1\right)}{(2\pi)^3}
\int \frac{M_B^*\; \theta(k_{F_B} - |\vec{k}|)}{\sqrt{k^2 + (M_B^*)^2}} \; d^3k ,
\ee
has an extra factor, denoted by
\be \label{eq:Csigma}
C(\bra\s\ket) = \left[w_B^\s - \tilde{w}_B^\s d g_{\s N}\bra\s\ket\right].
\ee
Note that the $d$ term (the scalar polarizability) in $C(\bra\s\ket)$
does not have the factor of $\reci{2}$ that is found
Eq.~(\ref{eq:effMQMC}), because of the differentiation.\par

Given this new term in the equation for the mean scalar field, we can
see that this allows feedback of the scalar field which is modelling
the internal degrees of freedom of the baryons.  This feedback
prevents certain values of $\bra\s\ket$ from being accessed.\par

{\bf 3.} \quad The couplings to the proton are re-determined by the
fit to saturation properties (minimum binding energy per baryon and
saturation density) with the new effective masses for the proton and
neutron. The couplings for QMC which provide a fit to saturated
nuclear matter are shown in Table~\ref{tab:QMCparams}.\par

Given these changes alone, QHD is transformed into QMC.  When we
compare the results of Section~\ref{sec:results} with those of
Ref.~\cite{RikovskaStone:2006ta} minor differences arise because the
QMC calculations in Ref.~\cite{RikovskaStone:2006ta} are performed at
Hartree-Fock level, whereas here they have been performed at Hartree
level (mean-field) only.\par

\begin{table}[!h]
\centering
\caption{\protect\label{tab:QMCparams} Couplings for QMC with the
  octet of baryons, fit to saturation of nuclear matter.\vspace{2mm}}
\begin{ruledtabular}
%\begin{tabular*}{\hsize}{@{\extracolsep{\fill}}ccccc}
\begin{tabular}{ccccc}
%hline
%hline
\phantom{space} & $g_{\s N}$ & $g_{\w N}$ & $g_{\rho}$ & \phantom{space} \\
& 8.278 & 8.417 & 8.333 & \\
%hline
%hline
\end{tabular}
\end{ruledtabular}
\end{table}

\section{Deconfined Quark Matter} \label{sec:quarkmatter}

We consider two models for a deconfined quark matter phase, both of
which model free quarks in $\beta$-equilibrium.  The first model, the
MIT bag model~\cite{Chodos:1974je}, is commonly used to describe the
quark matter phase because of its simplicity.\par

In this model we consider three quarks with fixed masses to possess
chemical potentials related to the independent chemical potentials of
Eq.~(\ref{eq:chempotrel}) via
\be \label{eq:qchempot}
\mu_u = \reci{3}\mu_n - \frac{2}{3}\mu_e,\qquad
\mu_d = \reci{3}\mu_n + \frac{1}{3}\mu_e,\qquad
\mu_s = \mu_d,
\ee
where quarks have a baryon charge of $\reci{3}$ since baryons contain
3 quarks.  Because the quarks are taken to be free, the chemical
potential has no vector interaction terms, and thus
\be \label{eq:quarkmu}
\mu_q = \sqrt{k_{F_q}^2 + m_q^2}\ ; \quad q\in\{u,d,s\}.
\ee
The EoS can therefore be solved under the conditions of
Eq.~(\ref{eq:equilconds}).\par

As an alternative model for deconfined quark matter, we consider a
simplified Nambu-Jona-Lasinio (NJL) model~\cite{Nambu:1961tp}, in
which the quarks have dynamically generated masses, ranging from
constituent quark masses at low densities to current quark masses at
high densities.  The equation for a quark condensate at a given
density (and hence, $k_F$) in NJL is similar to the scalar field in
QHD/QMC, and is written as
\be \label{eq:gap}
\bra\bar{\psi}\psi\ket = -4\ {\cal N}_c \int 
\frac{1}{(2\pi^3)} \frac{M_q^*\; \theta(k_{F}-|\vec{k}|)\; 
\theta(\Lambda - k_F)}{\sqrt{k^2+(M_q^*)^2}}\; d^3k,
\ee
where $M_q^*$ denotes the $k_F$ dependent (hence, density dependent)
quark mass; ${\cal N}_c$ is the number of color degrees of freedom of
quarks; and $\Lambda$ is the momentum cutoff. This is
self-consistently calculated via
\be \label{eq:cqmandcurrent}
M_q^* = M_{\rm current} - G\bra\bar{\psi}\psi\ket ,
\ee
where $G$ is the coupling and $M_{\rm current}$ the current quark mass.\par

\begin{figure}[!b]
\centering
\includegraphics[angle=90,width=0.9\textwidth]{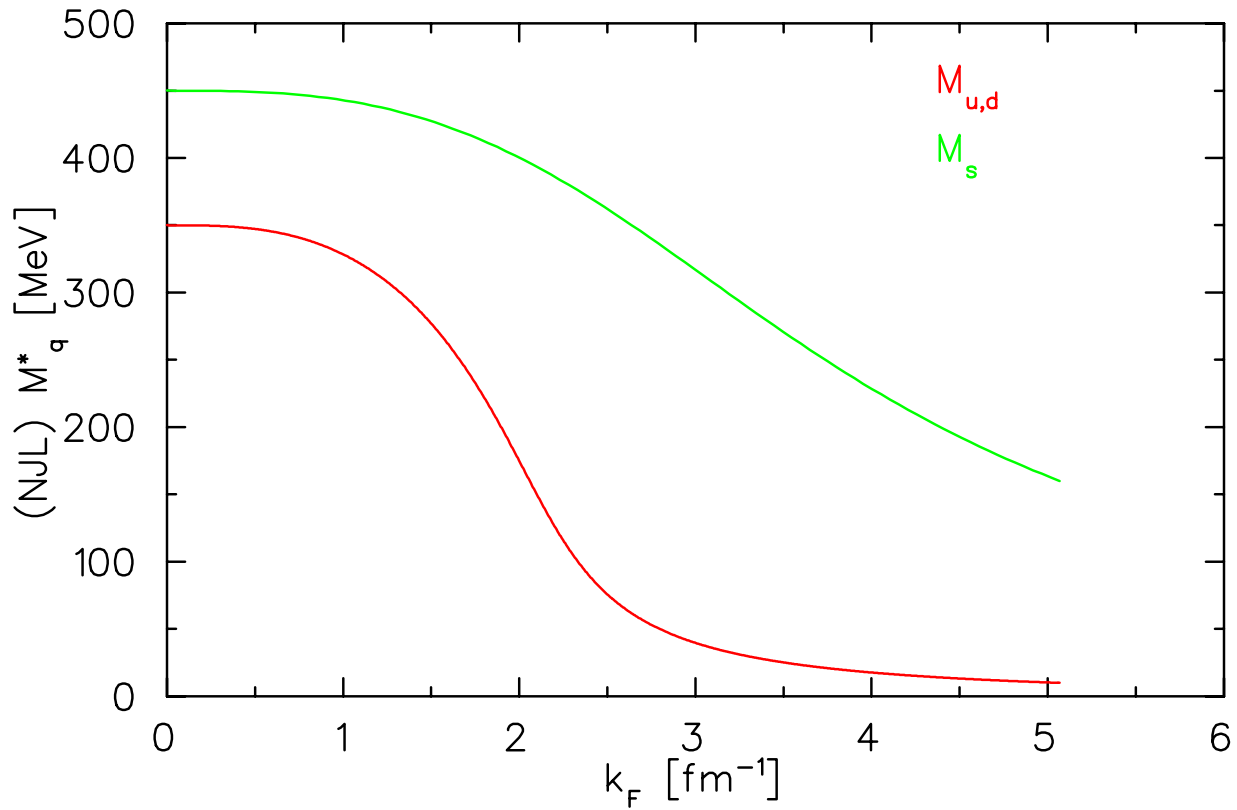}
\caption{(Color online) Density dependent (dynamic) masses for quarks
  using NJL.  The mass at $k_F = 0$ is the constituent quark mass, and
  the mass at the cutoff of $k_F = \Lambda$ is roughly the current
  quark mass.  This model successfully reproduces the behaviour found
  within the Schwinger-Dyson formalism, and we consider the model to
  be more sophisticated than the constant quark mass MIT bag model.
  \protect\label{fig:mkf}}
\end{figure}

To solve for the quark mass at each density, we must first find the
coupling, $G$, which yields the required constituent quark mass in
free space ($k_F=0$).  The coupling is assumed to remain constant as
the density rises. In free space, we can solve the above equations to
find the coupling
\be \label{eq:qcoupling}
G = \frac{(M_q^*-M_{\rm current})}{4\ {\cal N}_c }
\left[ \int 
\reci{(2\pi)^3}\frac{M_q^* \; \theta(|\vec{k}| - k_{F})\; 
\theta(\Lambda - k_F)}{\sqrt{k^2+(M_q^*)^2}}\; d^3k \right]^{-1} 
\Bigg|_{k_F = 0} \, .
\ee
We solve Eqs.~(\ref{eq:gap}--\ref{eq:qcoupling}) for ${\cal N}_c = 3$
to obtain constituent quark masses of \mbox{$M_{u,d} = 350~{\rm MeV}$}
using current quark masses of $M_{\rm current} = 10~{\rm MeV}$ for the
light quarks, and to obtain a constituent quark mass of \mbox{$M_s =
  450~{\rm MeV}$} using a current quark mass of $M_{\rm current} =
160~{\rm MeV}$ for the strange quark, with a momentum cutoff of
$\Lambda = 1~{\rm GeV}$.  At $k_F = 0$ we find the couplings to be
\be
\label{eq:NJLcouplings}
G_{u,d} = 0.148~{\rm fm}^2,\quad G_{s} = 0.105~{\rm fm}^2.
\ee
We can now use these parameters to evaluate the dynamic quark mass
$M_q^*$, for varying values of $k_F$, by solving Eq.~(\ref{eq:gap})
and Eq.~(\ref{eq:cqmandcurrent}) self-consistently.  The resulting
density dependence of $M_q^*$ is illustrated in Fig.~\ref{fig:mkf}.
This shows that the masses of the quarks eventually saturate and are
somewhat constant above a certain density.
We can then construct the EoS in the same way as we did for the MIT
bag model, but with density-dependent masses, rather than fixed
masses.\par

\section{Phase Transitions} \label{sec:PT}

\subsection{Equilibrium Conditions} \label{sec:equilconds}

We now have a description of hadronic matter with quark degrees of
freedom, but we are still faced with the issue that the baryons are
very densely packed.  We wish to know if it is more energetically
favourable for deconfined quark matter to be the dominant phase at a
certain density.  To do this, we need to find a point (if it exists)
at which stability is achieved between the hadronic phase and the
quark phase.\par

The condition for stability is that chemical, thermal, and mechanical
equilibrium between the hadronic ($H$) and quark ($Q$) phases is
achieved, and thus that the independent quantities in each phase are
separately equal. Thus the two independent chemical potentials,
($\mu_n,\mu_e$), are each separately equal to their counterparts in
the other phase, {\it i.e.}  $(\mu_n)_H=(\mu_n)_Q$, and
$(\mu_e)_H=(\mu_e)_Q$ (chemical equilibrium); the temperatures are
equal ($T_H=T_Q$) (thermal equilibrium); and the pressures are equal
($P_H = P_Q$) (mechanical equilibrium).  For a discussion of this
condition, see Ref.~\cite{Reif}.  We consider both phases to be cold
on the nuclear scale, and assume $T=0$, so the temperatures are by
construction equal.  We must therefore find the point at which, for a
given pair of independent chemical potentials, the pressures in both
the hadronic phase and the quark phase are the same.\par

To find the partial pressure of any baryon, quark, or lepton species,
$i$, we use
\be \label{eq:pressures}
P_i = \frac{\left(2J_B + 1\right) {\cal N}_c}{3(2\pi)^3}\int
\frac{k^2\; \theta(k_{F_i} - |\vec{k}|)}{\sqrt{k^2+(M_i^*)^2}}\; d^3k,
\ee
where ${\cal N}_c = 3$ for quarks, and ${\cal N}_c = 1$ for baryons
and leptons.  To find the total pressure in each phase we use
\be \label{eq:Hpressure}
P_H = \sum_B P_B  + \sum_\ell P_\ell 
+ \sum_{\alpha=\w,\rho} \reci{2} m_\alpha^2 \bra\alpha\ket^2
- \reci{2} m_\s^2 \bra\s\ket^2,
\ee
which is equivalent to Eq.~(\ref{eq:P_H}), and
\be \label{eq:Qpressure}
P_Q = \sum_q P_q + \sum_\ell P_\ell - B,
\ee
where $B$ in the quark pressure is the bag energy density. For the QMC
model described in Section~\ref{sec:QMC}, and a Fermi gas of quarks,
both with interactions with leptons for charge neutrality, a point
exists at which the condition of stability, as described above, is
satisfied.\par

At this point, it is equally favourable that hadronic matter and quark
matter are the dominant phase. Beyond this point, the quark pressure
is greater than the hadronic pressure, and so the quark phase has a
lower thermodynamic potential (through the relation $P=-\Omega$) and
the quark phase will be more energetically favourable.  To determine
the EoS beyond this point, we need to consider a mixed phase.\par

\begin{figure}[!b]
\centering
\includegraphics[width=0.9\textwidth]{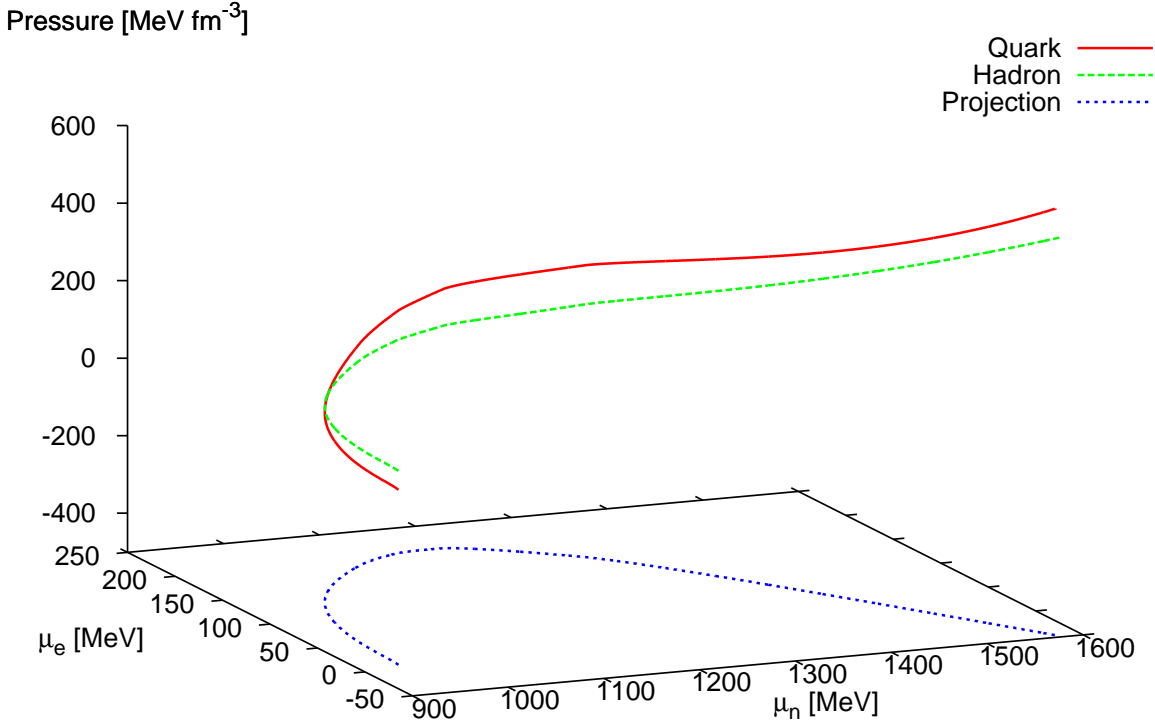}
\caption{(Color online) Illustrative locus of values for $\mu_e,
  \mu_n, P$ for phases of hadronic matter and deconfined quark matter.
  Note that pressure increases with density. and that a projection
  onto the $\mu_n\mu_e$ plane is a single line, as ensured by the
  chemical equilibrium condition.  \protect\label{fig:3d}}
\end{figure}

\subsection{Mixed Phase} \label{sec:MixedPhase}

We can model a mixed phase of hadronic and quark matter --- as opposed
to modelling a simple direct phase transition between the two, a
Maxwell construction, which would have a discontinuity in the density,
while retaining a constant pressure between the two phases --- using
the method of Glendenning.  A detailed description of this appears in
Ref.~\cite{Glendenning:2001pe}.\par

We solve for the hadronic EoS using the independent chemical
potentials as inputs for the quark matter EoS, as the order parameter,
$\rho$, the conserved baryon density, increases until we find a point
(if it exists) at which the pressure in the quark phase is equal to
that of the hadronic phase.  Once we have the density and pressure at
which the phase transition occurs, we change the order parameter from
the conserved baryon density to the quark fraction, $\chi$. If we
consider the mixed phase to be a fraction of the hadronic matter and a
fraction of the quark matter, then the mixed phase (MP) of matter will
have the following properties; the total density will be
\be 
\label{eq:mp_rho}
\rho_{\rm MP} = (1-\chi)\; \rho_{\rm HP} + \chi\; \rho_{\rm QP},
\ee
where $\rho_{\rm HP}$ and $\rho_{\rm QP}$ are the densities in the
hadronic and quark phases, respectively. The equivalent baryon density
in the quark phase,
\be 
\label{eq:equivrho}
\rho_{\rm QP} = \sum_q \rho_q = 3(\rho_u + \rho_d + \rho_s),
\ee
arises because of the restriction that a bag must contain 3
quarks.\par

According to the condition of mechanical equilibrium, the pressure in
the mixed phase will be
\be
\label{eq:mp_P}
P_{\rm MP} = P_{\rm HP} = P_{\rm QP}.
\ee
We can step through values $0 < \chi < 1$ and find the density at
which equilibrium is achieved, keeping the mechanical stability
conditions as they were above.  In the mixed phase we need to alter
our definition of charge neutrality; it becomes possible now that one
phase is (locally) charged, while the other phase carries the opposite
charge, making the system globally charge neutral.  This is achieved
by enforcing
\be \label{eq:mp_charge}
0 = (1-\chi)\; \rho^c_{\rm HP} + \chi\; \rho^c_{\rm QP} + 
\rho^c_{\ell} \, ,
\ee
where this time we are considering charge densities, which are simply
charge proportions of density and $\rho^c_{\ell}$ is the lepton charge
density. For example, the charge density in the quark phase is given
by
\be \label{eq:qp_charge}
\rho^c_{\rm QP} = \sum_q Q_q \rho_q = \frac{2}{3}\rho_u - \reci{3}\rho_d - \reci{3}\rho_s.
\ee
We continue to calculate the densities until we reach $\chi = 1$, at
which point the mixed phase is now entirely charge neutral quark
matter.  After this point, we continue with the EoS for pure charge
neutral quark matter, using $\rho$ as the order parameter.\par

\section{Stellar Solutions} \label{sec:TOV}

To test the predictions of these models, we find solutions of the
Tolman-Oppenheimer-Volkoff (TOV)~\cite{Oppenheimer:1939ne} equation
\be \label{eq:TOVdpdr}
\frac{dP}{dR} = 
-\frac{G\left(P + {\cal E}\right)\left(M(R)+4\pi R^3P\right)}{R(R-2GM(R))},
\ee
where the mass, $M(R)$, contained within a radius $R$ is found by
integrating the energy density
\be 
\label{TOVm}
M(R) = \int_0^R 4\pi r^2 {\cal E} \; dr,
\ee
and ${\cal E}$ and $P$ are the energy density and pressure in the EoS,
respectively.\par

Given an EoS and a choice for the central density of the star, this
provides static, spherically symmetric, non-rotating, gravitationally
stable stellar solutions for the total mass and radius of a star. For
studies of the effect of rapid rotation in General Relativity we refer
to Refs.~\cite{Lattimer:2004nj,Owen:2005fn}.  This becomes important
for comparison to experimental data, as only data for stellar masses
exists (with the single, disputed exception from \cite{Ozel:2006bv}),
we can use the model to predict the radii of the observed stars.\par

\section{Results} \label{sec:results}

To obtain numerical results, we solve the meson field equations,
Eqs.~(\ref{eq:MFsigma}--\ref{eq:MFrho}), with the conditions of charge
neutrality, fixed baryon density, and the equivalence of chemical
potentials given by Eq.~(\ref{eq:equilconds}), for various models.
Having found the EoS by evaluating the energy density,
Eq.~(\ref{eq:E_H}), and pressure, Eq.~(\ref{eq:P_H}), we can solve for
stellar solutions for an EoS using the TOV equation.  The radius of
the star is defined as the radius at which the pressure is zero and is
calculated using a fourth-order Runge-Kutta integration method.\par

\begin{figure}[!t]
\centering
\includegraphics[angle=90,width=0.9\textwidth]{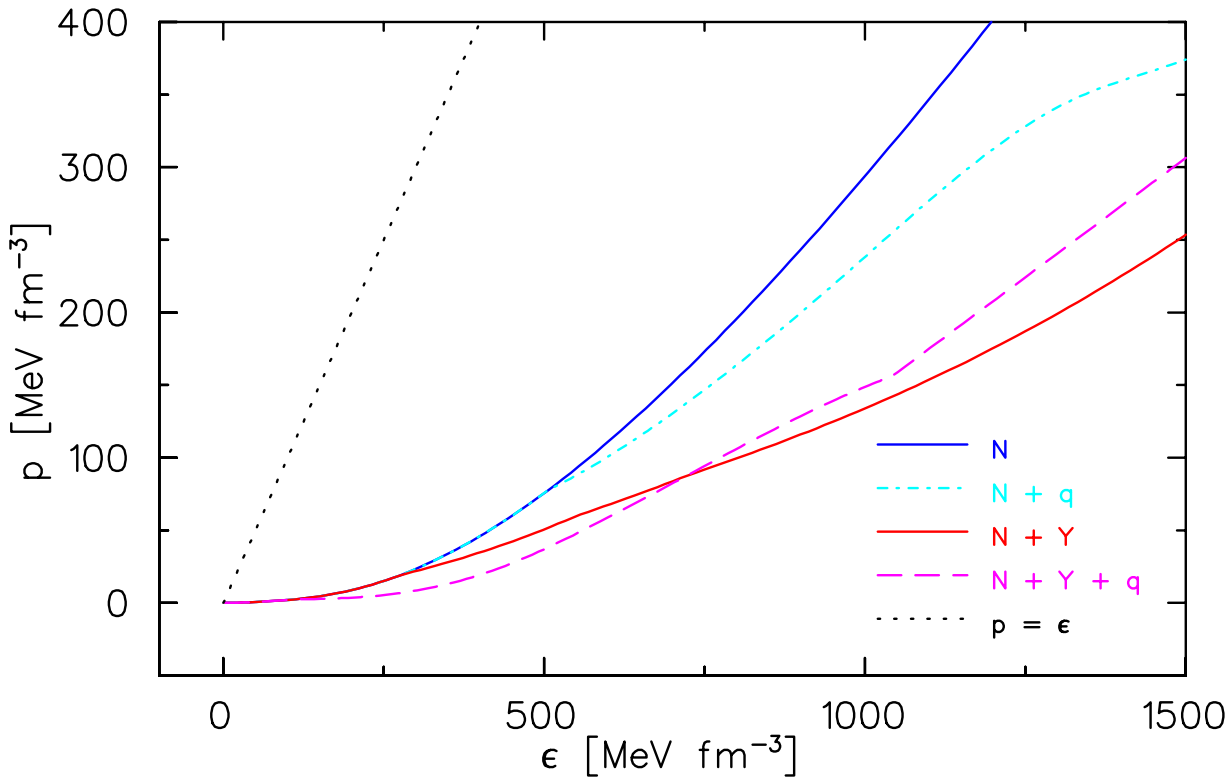}
\caption{(Color online) Equation of State for; nucleonic `N' matter
  modelled with octet QMC but where hyperons are explicitly forbidden;
  nucleonic matter where a phase transition to NJL modelled quark
  matter is permitted; baryonic `N+Y' matter modelled with octet QMC
  including hyperons; and baryonic matter where a phase transition to
  MIT bag modelled quark matter is permitted.  The line $P = {\cal E}$
  represents the causal limit, $v_{\rm sound} = c$.  The bends in
  these curves indicate a change in the composition of the EoS, such
  as the creation of hyperons or a transition to a mixed or quark
  phase.  Note that at low energies (densities) the curves are
  identical, where only nucleonic matter in $\beta$-equilibrium is
  present.  \protect\label{fig:EOSlots}}
\end{figure}

The EoS for octet QMC hadronic matter is shown in
Fig.~\ref{fig:EOSlots} alongside the same model when including a phase
transition to 3-flavor quark matter modelled with the MIT bag model,
and the results do not appear to differ much at this scale.  The
theoretical causality limit of $P = {\cal E}$ is also shown
(corresponding to the limit $v_{\rm sound} = c$) and we can see that
these models do not approach this limit at the scale displayed. This
is because of the softening of the EoS that occurs with the
introduction of hyperons, enlarging the Fermi sea to be filled and
reducing the overall pressure.\par

\begin{figure}[!t]
\centering
\includegraphics[angle=90,width=0.9\textwidth]{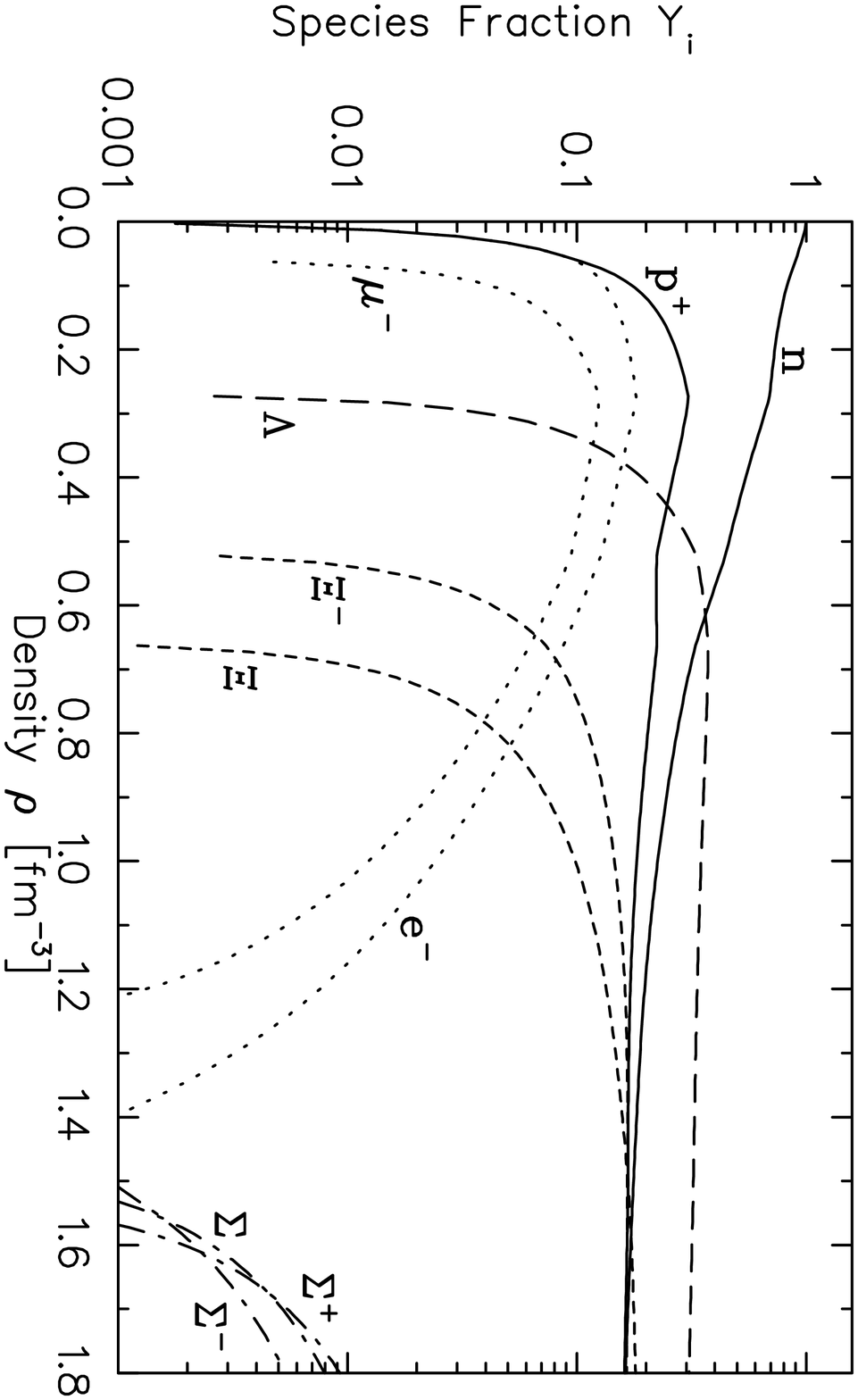}
\caption{Species fractions, $Y_i$, for octet QMC where a transition to
  a mixed phase is explicitly forbidden.  Note that in this case, all
  of the octet baryons contribute at some density, and that with
  increasing density the species fractions of $\Sigma$ hyperons are
  suppressed while the $\Lambda$ species fraction is
  enhanced. Parameters used here are shown in
  Table~\ref{tab:QMCparams}.  \protect\label{fig:specfrac_QMC}}
\end{figure}

The species fraction for each particle, $Y_i$, is simply the density
fraction of that particle, and is calculated via
\be 
\label{eq:Y}
Y_i = \frac{\rho_i}{\rho}\ ;\quad i\in \{ p,n,\Lambda,\Sigma^+,\Sigma^0,\Sigma^-,\Xi^0,\Xi^-,\ell,q \} \, ,
\ee
where $\rho$ is the total baryon density. The species fractions for
octet QMC when a phase transition is neglected are shown in
Fig.~\ref{fig:specfrac_QMC}, where we note that the $\Lambda$ species
fraction is enhanced and the $\Sigma$ species fractions are suppressed
with increasing density.  From the investigations by Rikovska-Stone
{\it et al.}~\cite{RikovskaStone:2006ta} we expect that the $\Sigma$
would disappear entirely if we were to include Fock terms.\par

\begin{figure}[!t]
\centering
\includegraphics[angle=90,width=0.9\textwidth]{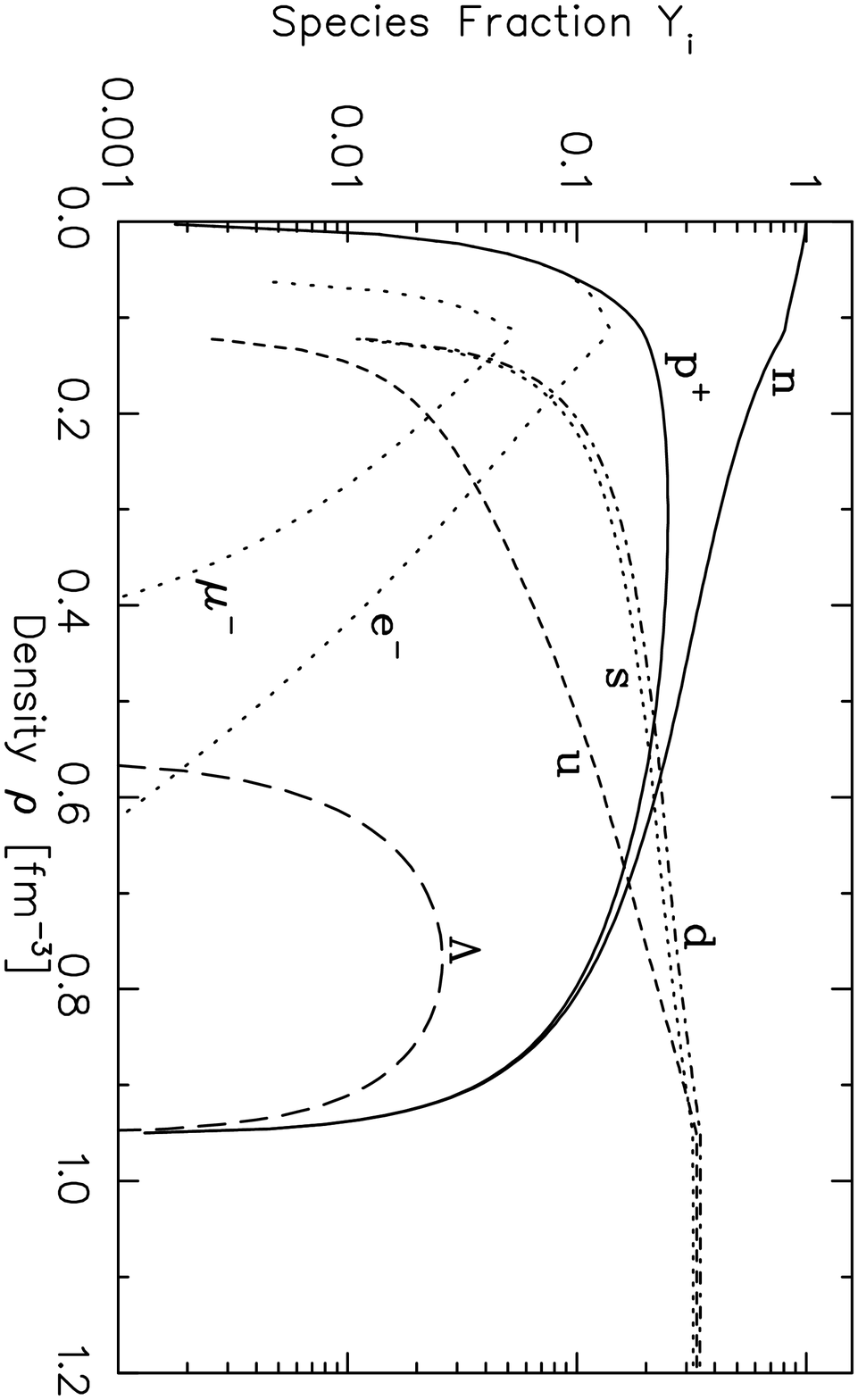}
\caption{Species fractions, $Y_i$, for octet QMC (the same as in
  Fig.~\ref{fig:specfrac_QMC}) but where now we allow the phase
  transition to a mixed phase involving quark matter modelled with the
  MIT bag model, and subsequently to a pure deconfined quark matter
  phase. Parameters used here are summarised in
  Table~\ref{tab:results}. Note that with these parameters, the
  $\Lambda$ is the only hyperon to appear in the mixed phase, and does
  so at a much higher density than the case where the transition to a
  mixed phase is forbidden. We also note that with these parameters,
  the transition to a mixed phase occurs below saturation density,
  $\rho_0$.  \protect\label{fig:SpecFrac_PTQMC}}
\end{figure}

The value of compression modulus and effective nucleon mass at
saturation are frequently used as a comparison to experimental
evidence. Models which neglect quark-level interactions, such as QHD,
typically predict much higher values for the compression modulus than
experiments suggest. In the symmetric (nuclear) matter QHD model
described in this paper, we find values of $(M^*/M)_{\rm sat} = 0.56$
and $K = 525~{\rm MeV}$ which are in agreement
with~\cite{Serot:1984ey}, but as stated in that reference, not with
experiment. For QMC we find a significant improvement in the
compression modulus; $K = 280~{\rm MeV}$ which lies at the upper end
of the experimental range. The nucleon effective mass at saturation
for QMC is found to be $(M^*)_{\rm sat} = 735~{\rm MeV}$, producing
$(M^*/M)_{\rm sat} = 0.78$.\par

When we calculate the EoS including a mixed phase and subsequent pure
quark phase, we find that small changes in the parameters can
sometimes lead to very significant changes.  In particular, the bag
energy density, $B$, and the quark masses in the MIT bag model have
the ability to both move the phase transition points, and to vary the
constituents of the mixed phase.  We have investigated the range of
parameters which yield a transition to a mixed phase and these are
summarised in Table~\ref{tab:results}.  For illustrative purposes we
show an example of species fractions for a reasonable set of
parameters ($B^{1/4}=180~{\rm MeV}$ and $m_{\rm u,d,s} = 3,7,95~{\rm
  MeV}$) in Fig.~\ref{fig:SpecFrac_PTQMC}. Note that in this case the
$\Lambda$ hyperon enters the mixed phase briefly (and at a low species
fraction).\par

Note that the transition density of $\rho_{\rm MP} \sim 0.12~{\rm
  fm}^{-3}$ produced by the combination of the octet QMC and MIT bag
models (as shown in Fig.~\ref{fig:SpecFrac_PTQMC}) is clearly not
physical as it implies the presence of deconfined quarks at densities
less than $\rho_0$.\par

\begin{figure}[!t]
\centering
\includegraphics[angle=90,width=0.9\textwidth]{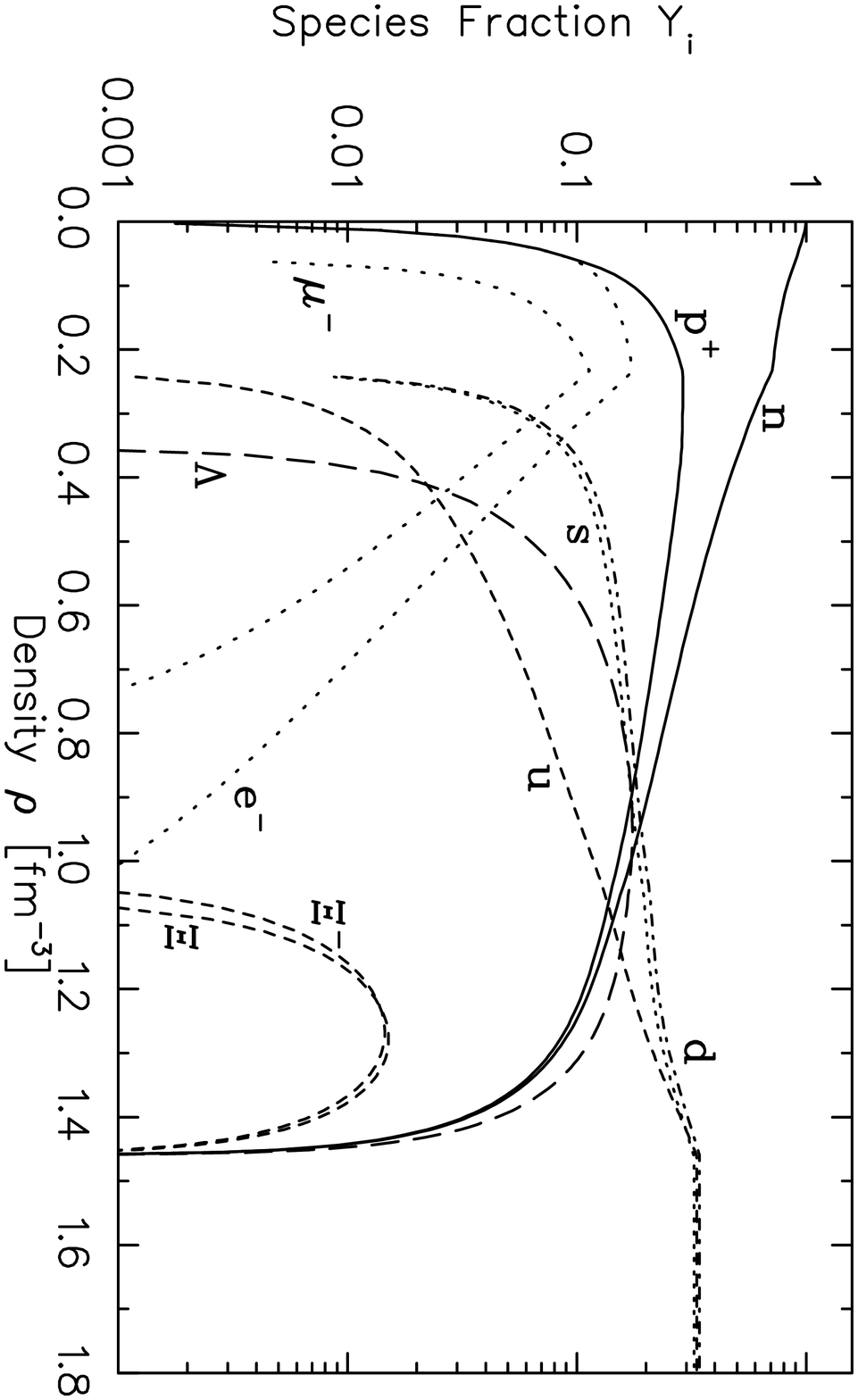}
\caption{Species fractions, $Y_i$, for octet QMC (the same as in
  Fig.~\ref{fig:SpecFrac_PTQMC} but now where the bag energy density
  has been increased to $B^{1/4}=195~{\rm MeV}$).  Note that now the
  appearance of hyperons occurs at a smaller density than in the case
  of Fig.~\ref{fig:SpecFrac_PTQMC}, the transition to a mixed phase
  occurs at a slightly larger density, and that now $\Xi$ hyperons are
  present in the mixed phase.  \protect\label{fig:SpecFrac_PTQMC_195}}
\end{figure}

With small changes to parameters, such as those used to produce
Fig.~\ref{fig:SpecFrac_PTQMC_195} in which the bag energy density is
given a slightly higher value from that used in
Fig.~\ref{fig:SpecFrac_PTQMC} ($B^{1/4}$ increased from $180~{\rm
  MeV}$ to $195~{\rm MeV}$, but the quark masses remain the same), it
becomes possible for the $\Xi$ hyperons to also enter the mixed phase,
albeit in that case with small species fractions, $Y_\Sigma, Y_\Xi
\leq 0.02$.
%Note that with this small change to the parameters the density at which 
%a transition to a mixed phase is produced is increased significantly.\par

\begin{figure}[!t]
\centering
\includegraphics[angle=90,width=0.9\textwidth]{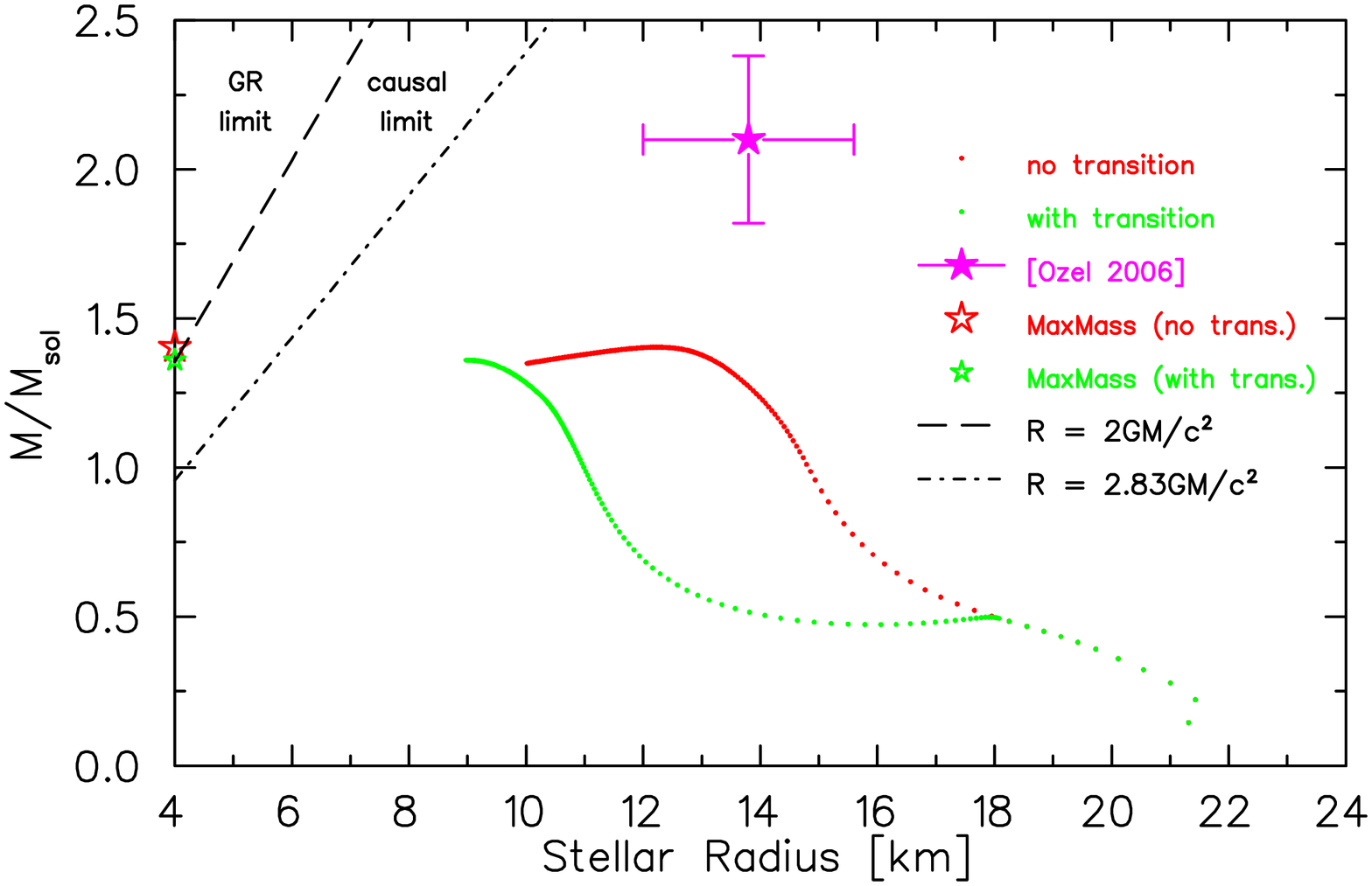}
\caption{(Color online) Solutions of the TOV equations for the total
  stellar mass and radius for octet QMC, where a phase transition to
  mixed phase is explicitly forbidden (as shown in
  Fig.~\ref{fig:specfrac_QMC}) and the same model with an allowed
  phase transition to 3-flavor quark matter modelled with the MIT bag
  model (as shown in Fig.~\ref{fig:SpecFrac_PTQMC}).  Also shown is
  the data point from~\cite{Ozel:2006bv}.  The points on the vertical
  axis are the maximum masses in the respective models. The causal and
  general relativistic limits on mass and radius are also shown.
  \protect\label{fig:TOV_QMC}}
\end{figure} 

\begin{figure}[!b]
\centering \includegraphics[angle=90,width=0.9\textwidth]{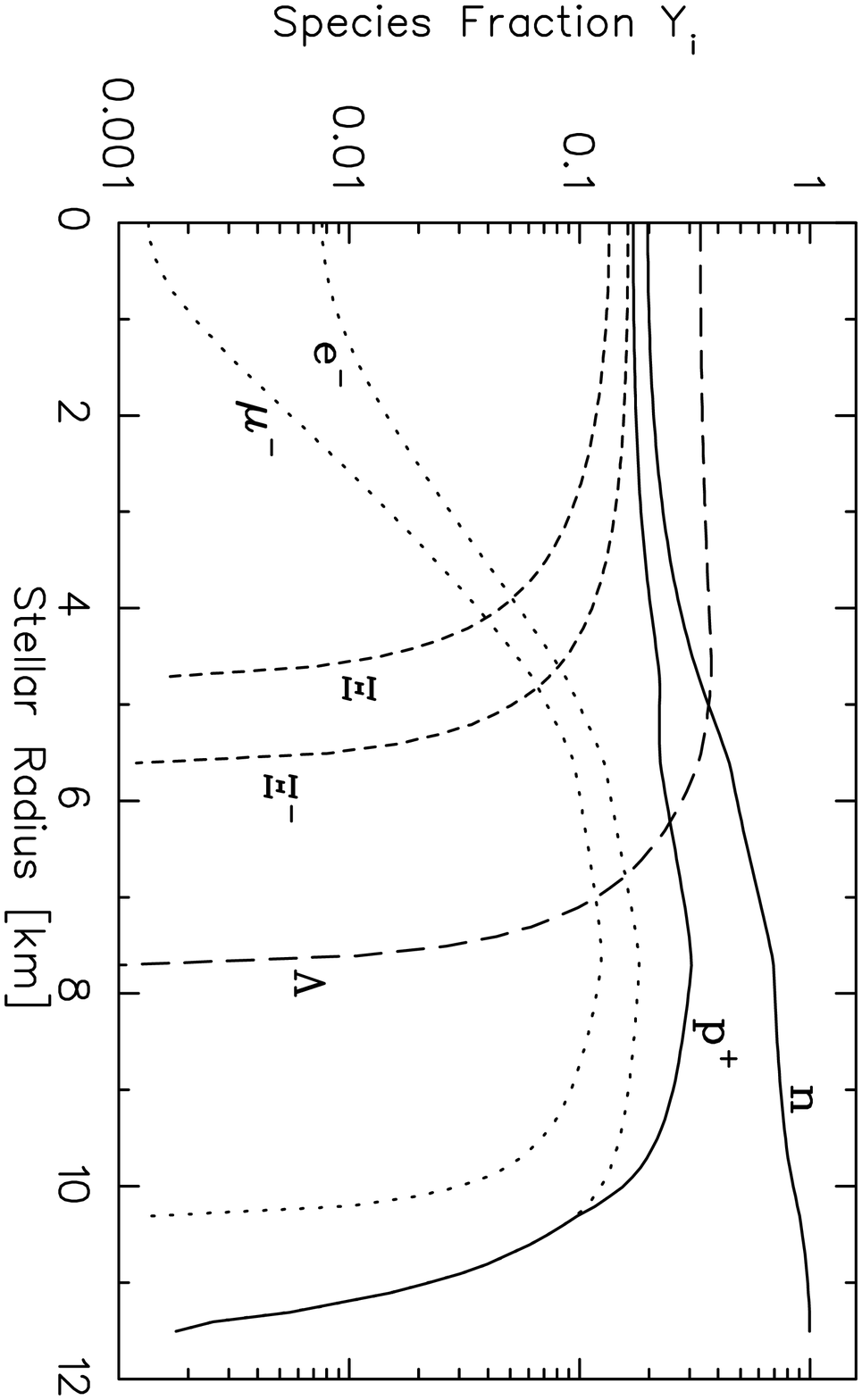}
\caption{Species fractions for octet QMC in $\beta$-equilibrium, where
  the phase transition to a mixed phase is explicitly forbidden, as a
  function of stellar radius for a stellar solution with a total mass
  of 1.2 $M_\odot$. The parameters used here are the same as those
  used to produce Fig.~\ref{fig:specfrac_QMC}.
  \protect\label{fig:TOVRAD_noPT}}
\end{figure}

\begin{figure}[!t]
\centering
\includegraphics[angle=90,width=0.9\textwidth]{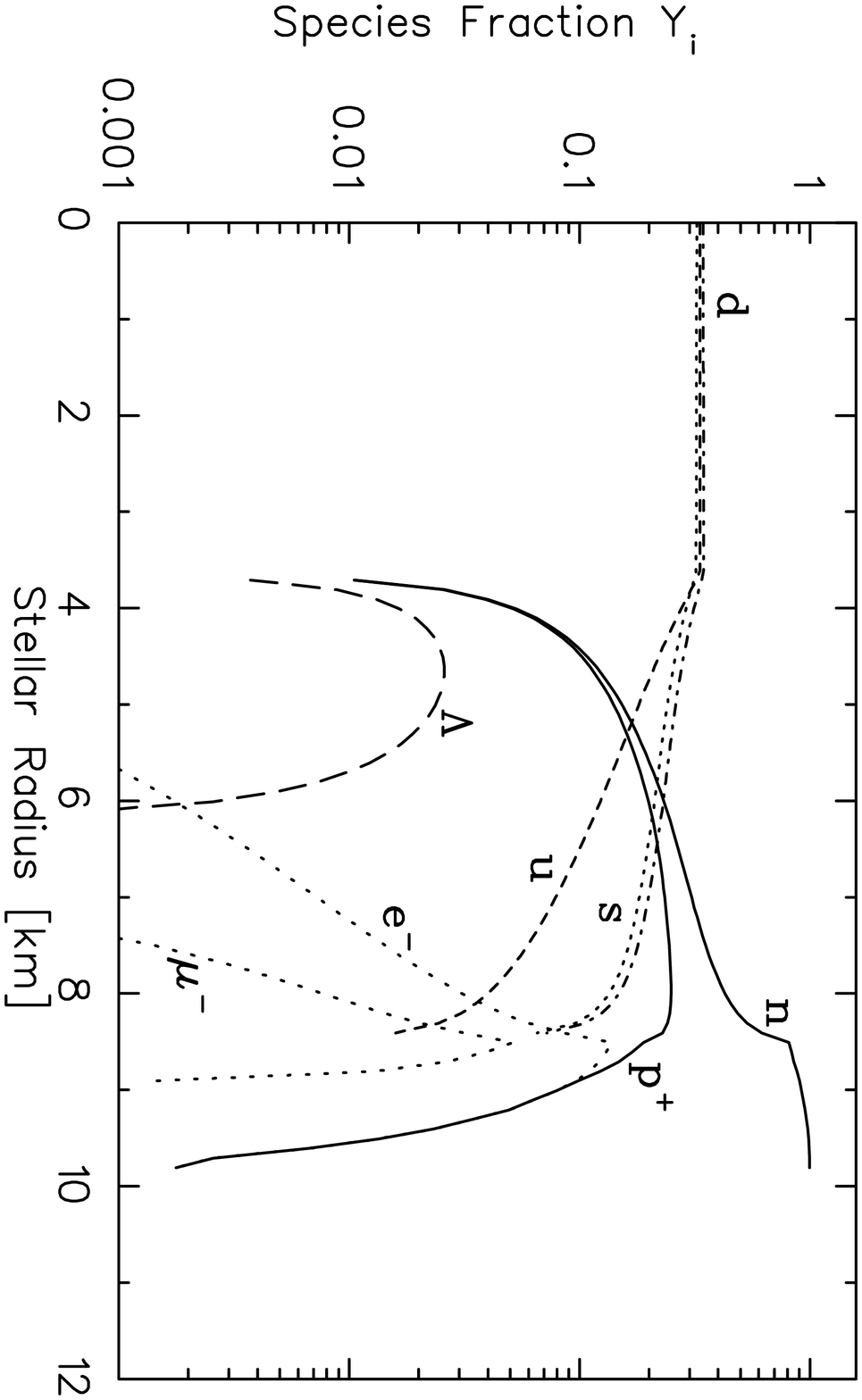}
\caption{Species fractions for octet QMC with a phase transition to
  3-flavor quark matter modelled with the MIT bag model, as a function
  of stellar radius for a stellar solution with a total mass of 1.2
  $M_\odot$.  The parameters used here are the same as those used to
  produce Fig.~\ref{fig:SpecFrac_PTQMC}.  Note that in this case one
  finds pure deconfined 3-flavor quark matter at the core (all of some
  3.5~km) of this star, and still a small proportion of $\Lambda$ in
  the mixed phase.  \protect\label{fig:TOVRAD_PT}}
\end{figure}

The TOV solutions for octet QMC with and without a phase transition to
a mixed phase are shown in Fig.~\ref{fig:TOV_QMC}. The stellar masses
produced using these methods are similar to observed neutron star
masses.  Once we have solved the TOV equations, we can examine
individual solutions and determine the species content for specific
stars.  If we examine the solutions with a stellar mass of
$M=1.2~M_\odot$, where $M_\odot$ is a solar mass, for the set of
parameters used to produce Figs.~\ref{fig:specfrac_QMC} and
\ref{fig:SpecFrac_PTQMC}, we can find the species fraction as a
function of stellar radius to obtain a cross-section of the star. This
is shown in Fig.~\ref{fig:TOVRAD_noPT} for the case of no phase
transition, and Fig.~\ref{fig:TOVRAD_PT} for the case where we allow a
transition to a mixed phase, and subsequently to a quark matter
phase.\par

\begin{figure}[!t]
\centering
\includegraphics[angle=90,width=0.9\textwidth]{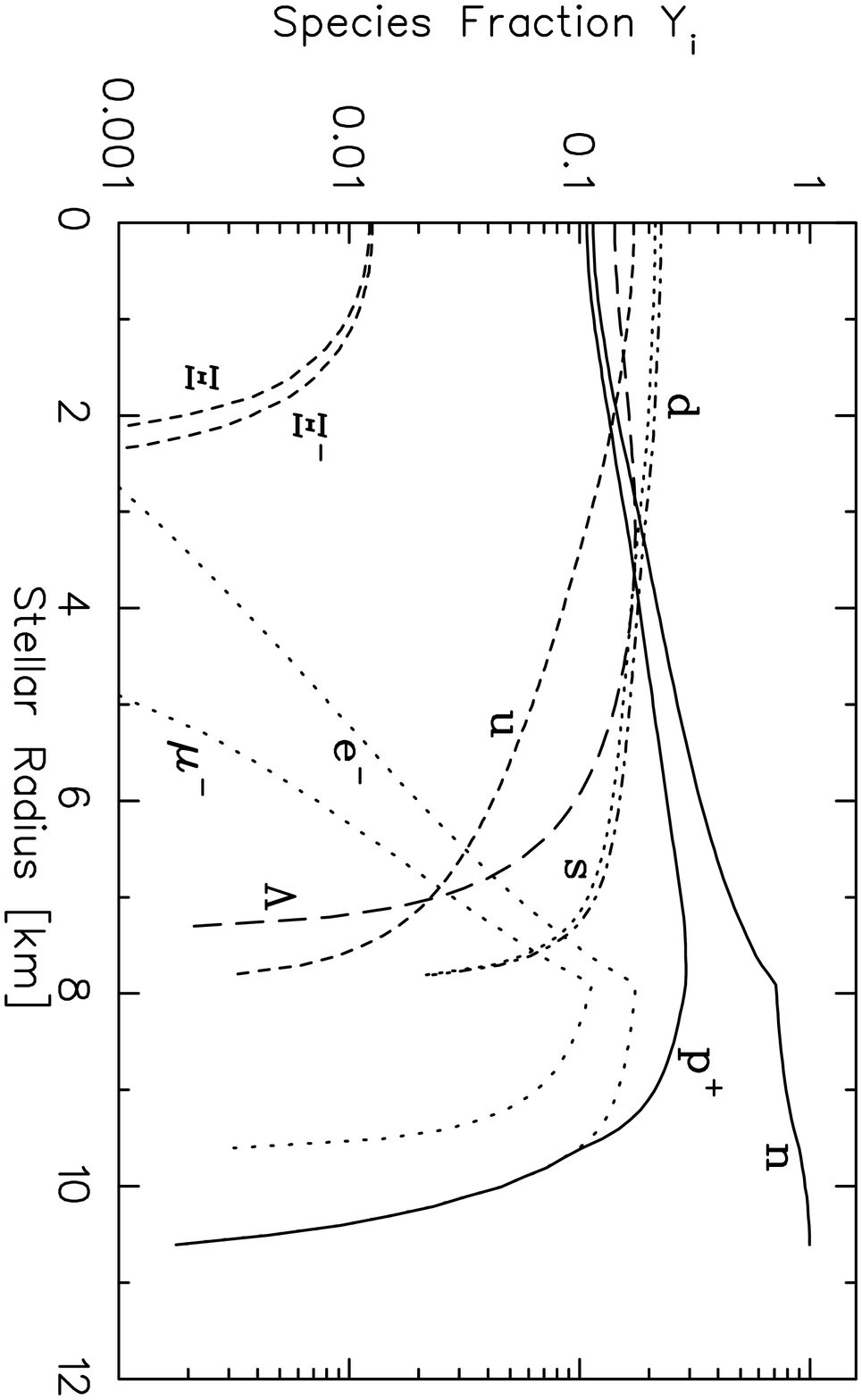}
\caption{Example of the interior of a star of total stellar mass $M =
  1.2 M_\odot$ where the bag energy density is given a slightly higher
  value from that used in Fig.~\ref{fig:TOVRAD_PT} (increased from
  $B^{1/4} = 180~{\rm MeV}$ to $195~{\rm MeV}$), but the quark masses
  remain the same. This illustrates that with relatively minor
  adjustments to the parameters, large changes can be introduced to
  the final solution.  In this case $\Xi$ hyperons can provide a
  nonzero contribution to the composition of a star.  Note that in
  this case, quark matter appears at 8~km, and at the core there
  exists a mixed phase containing nucleons, quarks, as well as
  $\Lambda$, $\Xi^0$, and $\Xi^-$ hyperons.
  \protect\label{fig:SpecFrac1.2_QMC}}
\end{figure}

If we now examine the stellar solution with mass $M=1.2~M_\odot$ of
the set of parameters ($B^{1/4}=195~{\rm MeV}$ and $m_{\rm u,d,s} =
3,7,95~{\rm MeV}$ used to produce Fig.~\ref{fig:SpecFrac_PTQMC_195})
as shown in Fig.~\ref{fig:SpecFrac1.2_QMC}, we note that the quark
content of this 10.5~km star reaches out to around 8~km, and that the
core of the star contains roughly equal proportions of
%around $1.5\% \ \Xi^-$, $0.6\% \ \Xi^0$, $11\% \ \Lambda$, and
%roughly $10\%$ of each of
protons, neutrons and $\Lambda$ hyperons with $Y_i \simeq 10\%$.\par

\begin{figure}[!t]
\centering
\includegraphics[angle=90,width=0.9\textwidth]{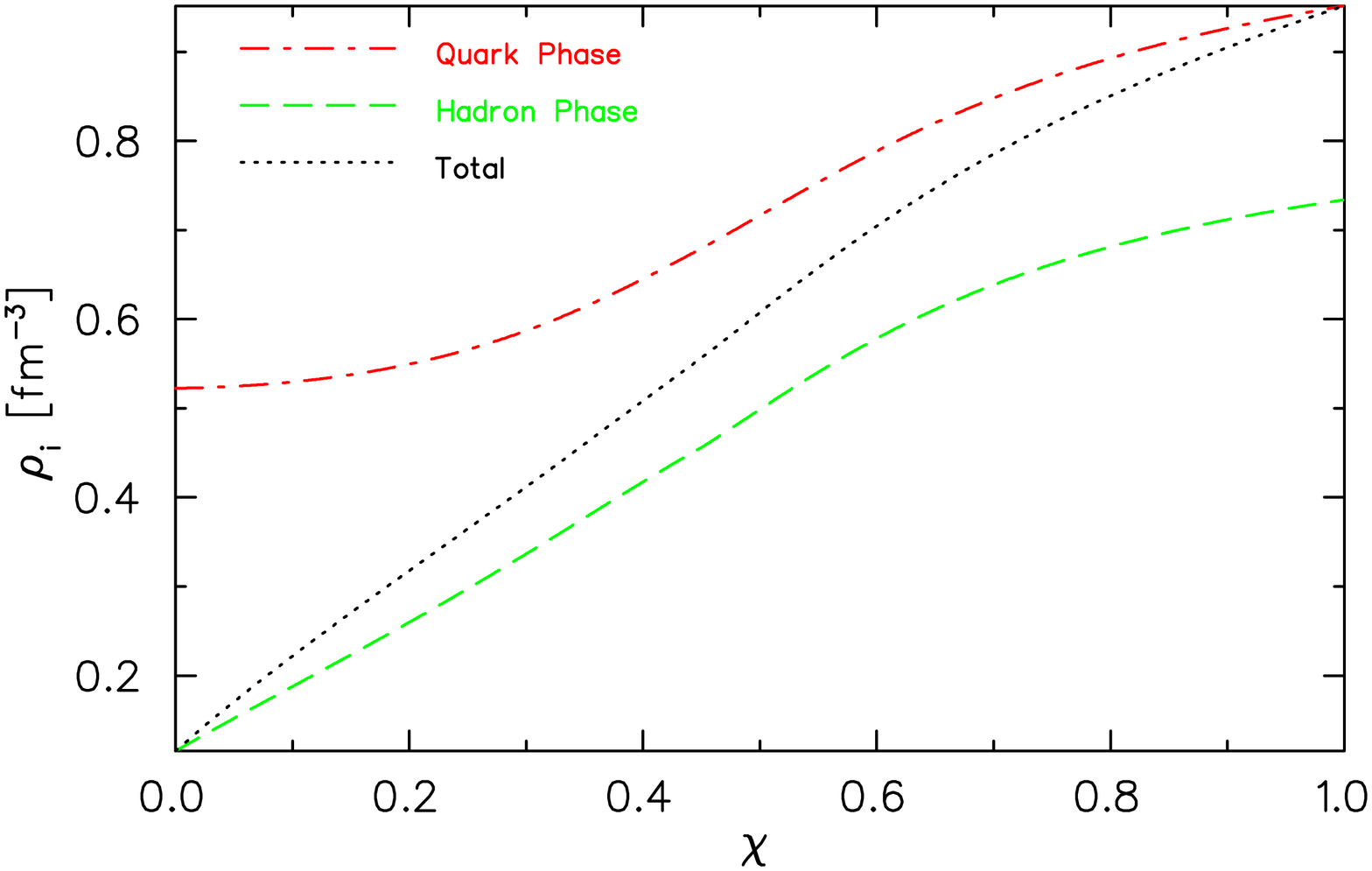}
\caption{(Color online) Densities in the mixed phase for octet QMC
  mixed with 3-flavor quark matter modelled with the MIT bag model.
  Note that at all values of $\chi$ (the mixing parameter according to
  Eq.~(\ref{eq:mp_rho})), the equivalent quark baryon density is
  greater than the hadronic baryon density, allowing the total baryon
  density to increase monotonically.  The total density is found via
  Eq.~(\ref{eq:mp_rho}).  \protect\label{fig:densities}}
\end{figure}

Within a mixed phase, we require that for a given pair of $\mu_n$ and
$\mu_e$ at any value of the mixing parameter $\chi$, the quark density
is greater than the hadronic density. This condition ensures that the
total baryon density increases monotonically within the range
$\rho_{\rm QP} > \rho_{\rm MP} > \rho_{\rm HP}$, as can be seen in
Eq.~(\ref{eq:mp_rho}). An example of this is illustrated in
Fig.~\ref{fig:densities} for a mixed phase of octet QMC and 3-flavor
quark matter modelled with the MIT bag model.\par

%Since the proportion of quark matter in the 
%mixed phase ranges from $0\%$ to 100$\%$, 
%satisfactory solutions require that for given $\mu_n$, $\mu_e$ and 
%$P_{\rm MP}$, the quark density is greater than the 
%hadronic density, otherwise the total baryon density 
%would decrease with increasing $\chi$. The respective 
%densities (within a mixed phase of octet QMC and 3-flavor 
%quark matter modelled with the MIT bag model) are shown in Fig.~\ref{fig:densities}, where we see  
%that the quark density is indeed greater than the hadronic density at 
%all $\chi$, and that the total baryon density 
%shifts from the hadronic curve to the quark curve as 
%$\chi$ increases.\par

The use of quark masses corresponding to the NJL model results in a
quark density that is lower than the hadronic density, and as a result
there are no solutions for a mixed phase in which the proportion of
quarks increases with fraction $\chi$, while at the same time the
total baryon density increases.  It may be possible that with smaller
constituent quark masses at low density, the Fermi momenta would
provide sufficiently high quark densities, but we feel that it would
be unphysical to use any smaller constituent quark masses. This result
implies that, at least for the model we have investigated, dynamical
chiral symmetry breaking (in the production of constituent quark
masses at low density) prevents a phase transition from a hadronic
phase to a mixed phase involving quarks.\par

\begin{figure}[!b]
\centering
\includegraphics[angle=90,width=0.9\textwidth]{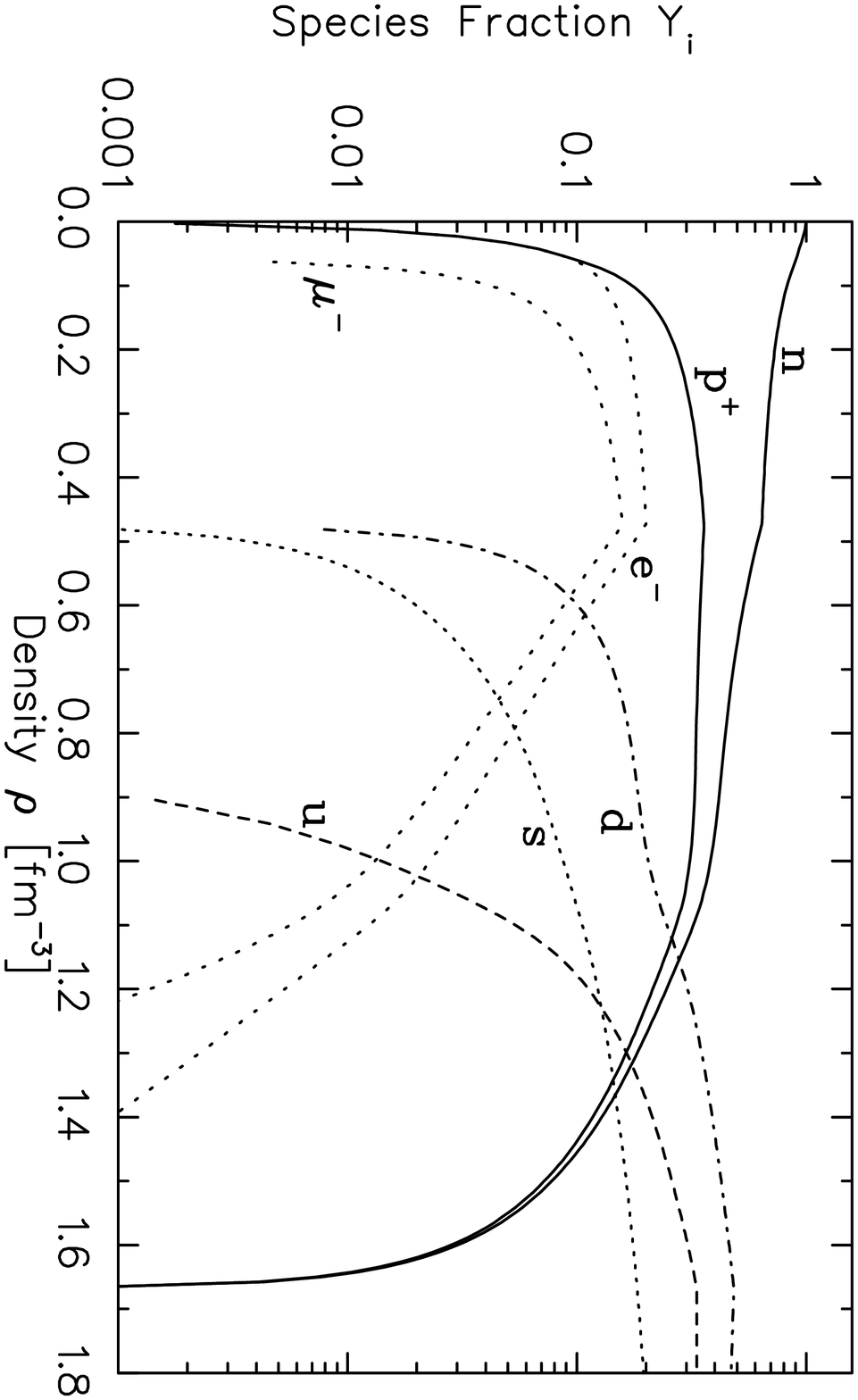}
\caption{Species fractions for a phase transition from QMC nuclear
  matter to 3-flavor quark matter modelled with NJL.  Note that in
  this unphysical case, a phase transition is possible, and occurs at
  a value of about $\rho = 0.47~{\rm fm}^{-3}$. We note the
  coincidence with the density of one baryon per baryon volume, but
  draw no conclusions from this. A similar transition from QMC nuclear
  matter to 3-flavor quark matter modelled with the MIT bag model
  produces results almost identical to those of
  Fig.~\ref{fig:SpecFrac_PTQMC} except that in that case there is no
  contribution from the $\Lambda$ hyperon.
  \protect\label{fig:QMCnuclear}}
\end{figure}

We do note, however, that if we restrict consideration to nucleons
only within the QMC model (with the same parameters as octet QMC), and
represent quark matter with the NJL model, we do in fact find a
possible mixed phase. More surprisingly, the phase transition density
for this combination is significantly larger than the case where
hyperons are present. An example of this is shown in
Fig.~\ref{fig:QMCnuclear} with parameters found in
Table~\ref{tab:results}.  This produces a mixed phase at about $3
\rho_0$ ($\rho = 0.47~{\rm fm}^{-3}$) and a pure quark matter phase
above about $10.5 \rho_0$ ($\rho = 1.67~{\rm fm}^{-3}$). We note the
coincidence of this phase transition density with the density
corresponding to one nucleon per nucleon volume, with the
aforementioned assumption of $R^{\rm free}_N = 0.8~{\rm fm}$, though
we do not draw any conclusions from this.  Performing this calculation
with quark matter modelled with the MIT bag model produces results
similar to those of Fig.~\ref{fig:SpecFrac_PTQMC} except of course
lacking the $\Lambda$ hyperon contribution. Although this example does
show a phase transition, the omission of hyperons is certainly
unrealistic. This does however illustrate the importance and
significance of including hyperons, in that their inclusion alters the
chemical potentials which satisfy the equilibrium conditions in such a
way that the mixed phase is no longer produced.\par

\begin{figure}[!t]
\centering
\includegraphics[angle=90,width=0.9\textwidth]{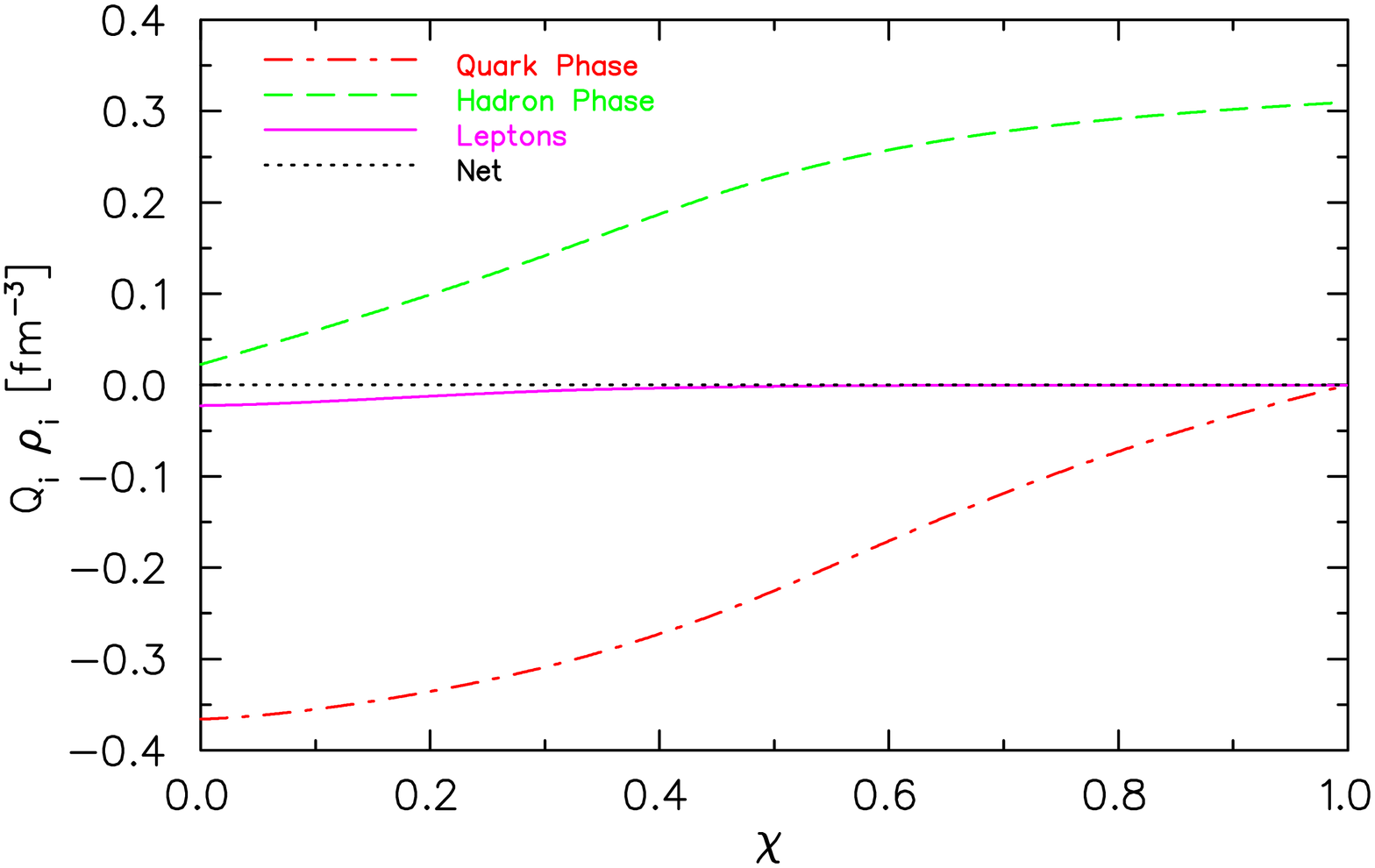}
\caption{(Color online) Charge densities (in units of the proton
  charge per cubic fm) in the mixed phase for a transition from octet
  QMC to 3-flavor quark matter modelled with the MIT bag model. Note
  that following the mixed phase, the quarks are able to satisfy
  charge neutrality with no leptons.  $\chi$ is the mixing parameter
  within the mixed phase according to Eq.~(\ref{eq:mp_rho}).
  \protect\label{fig:chargedensities}}
\end{figure}

For each of the cases where we find a phase transition from baryonic
matter to quark matter, the solution consists of negatively charged
quark matter, positively charged hadronic matter, and a small
proportion of leptons, to produce globally charge neutral matter.  The
proportions of hadronic, leptonic and quark matter throughout the
mixed phase (for example, during a transition from octet QMC matter to
3-flavor quark matter modelled with the MIT bag model) are displayed
in Fig.~\ref{fig:chargedensities}.  A summary of the results of
interest is given in Table~\ref{tab:results}.\par

\begin{table}[!b]
\centering
\caption{\protect\label{tab:results}Table of species content ($N$ =
  nucleons, $Y$ = hyperons, $\ell$ = leptons, $q$ = quarks); inputs
  ($B^{1/4}$, $m_q$); and results for octet QMC and quark models
  presented in this paper.  $\rho_{\rm Y}$, $\rho_{\rm MP}$ and
  $\rho_{\rm QP}$ represent the density at which hyperons first appear
  ($\Lambda$ is invariably the first to enter in these calculations);
  the density at which the mixed phase begins; and the density at
  which the quark phase begins, respectively.  Figures for selected
  parameter sets are referenced in the final column.  Dynamic NJL
  quark masses are determined by
  Eqs.~(\ref{eq:gap}--\ref{eq:qcoupling}).
\vspace{2mm}}
\begin{ruledtabular}
%begin{tabular*}{\hsize}{@{\extracolsep{\fill}}|l|cc|ccc|l|}
\begin{tabular}{lllllll}
%\begin{tabular}{lclcccl}
%\hline
%\hline
Particles: & 
$B^{1/4}~({\rm MeV})$ & 
$\{m_u,m_d,m_s\}~({\rm MeV})$ & 
$\rho_{\rm Y}~({\rm fm}^{-3})$ & 
$\rho_{\rm MP}~({\rm fm}^{-3})$ & 
$\rho_{\rm QP}~({\rm fm}^{-3})$ &Figure:\\
&&&&&& \\[-3mm]
%&\multicolumn{2}{c}{(MeV)}&\multicolumn{3}{c}{(${\rm fm}^{-3}$)}&\\[1mm]
\hline
&&&&&& \\[-2mm]
N, Y, $\ell$   & --- & ---                & 0.27 & ---  & ---  & Fig.~\ref{fig:specfrac_QMC} \\
N, Y, $\ell$, q & 180 & \{3, 7, 95\}      & 0.55 & 0.12 & 0.95 & Fig.~\ref{fig:SpecFrac_PTQMC} \\
N, Y, $\ell$, q & 195 & \{3, 7, 95\}      & 0.35 & 0.24 & 1.46 & Fig.~\ref{fig:SpecFrac_PTQMC_195} \\
N, Y, $\ell$, q & 170 & \{30, 70, 150\}   & 0.56 & 0.10 & 0.87 & --- \\
N, Y, $\ell$, q & 175 & \{100, 100, 150\} & 0.44 & 0.16 & 1.41 & --- \\
N, $\ell$, q    & 180 & Dynamic (NJL)     & ---  & 0.47 & 1.67 & Fig.~\ref{fig:QMCnuclear} 
%\hline
%\hline
\end{tabular}
\end{ruledtabular}
\end{table}

Results for larger quark masses are not shown, as they require a much
lower bag energy density to satisfy the equilibrium conditions.  For
constituent quark masses, we find that no phase transition is possible
for any value of the bag energy density, as the quark pressure does
not rise sufficiently fast to overcome the hadronic pressure. This is
merely because the mass of the quarks does not allow a sufficiently
large Fermi momentum at a given chemical potential, according to
Eq.~(\ref{eq:quarkmu}).\par

\section{Conclusions} \label{sec:conc}

We have produced several EoS that simulate a phase transition from
octet QMC modelled hadronic matter, via a continuous Glendenning style
mixed phase to a pure, deconfined quark matter phase. This should
correspond to a reasonable description of the relevant degrees of
freedom in each density region.  The models used here for quark matter
provide a framework for exploring the way that this form of matter may
behave, in particular under extreme conditions.  The success of the
QMC model in reproducing a broad range of experimental data gives us
considerable confidence in this aspect of these calculations, and
provides a reasonable hadronic sector and calculation framework, which
then awaits improvement in the quark sector to produce realistic
stellar solutions.\par

We have presented EoS and stellar solutions for octet QMC matter at
Hartree level. We have explored several possible phase transitions
from this hadronic sector to a mixed phase involving 3-flavor quark
matter. The corresponding EoS demonstrate the complexity and intricacy
of the solutions as well as the dependence on small changes in
parameters. The stellar solutions provide overlap with the lower end
of the experimentally acceptable range.\par

Several investigations were made of the response of the model to a
more sophisticated treatment of the quark masses in-medium, namely the
NJL model. In that model the quark masses arise from dynamical chiral
symmetry breaking and thus take values typical of constituent quarks
at low density and drop to current quark masses at higher
densities. The result is that no transition to a mixed phase is
possible in this case.\par

The omission of hyperons in the QMC model yields a transition to a
mixed phase of either NJL or MIT bag model quark matter, as the
hadronic EoS is no longer as soft.  This observation makes it clear
that hyperons can play a significant role in the EoS. However, we
acknowledge that their presence in neutron stars remains
speculative.\par

The models considered here reveal some important things about the
possible nature of the dense nuclear matter in a neutron star.  It
seems that if dynamical chiral symmetry does indeed result in typical
constituent quark masses in low density quark matter, then a phase
transition from hadronic matter to quark matter is unlikely. This
result invites further investigation.\par

The results presented in Fig.~\ref{fig:TOV_QMC} indicate that the
model in its current form is unable to reproduce sufficiently massive
neutron stars to account for all observations, notably the observed
stellar masses of $1.45~M_\odot$ and larger. This is a direct result
of the softness of the EoS. This issue will be explored in a future
publication via the inclusion of Fock terms, which have been shown to
have an effect on the scalar and vector
potentials~\cite{Krein:1998vc}.\par

Many open questions remain to be investigated in further work,
including the effects of Fock terms, and the density dependence of the
bag energy density in the quark phase, which can be calculated
explicitly within the NJL model. The quark matter models used here are
still not the most sophisticated models available, and further work
may involve an investigation of the effects of color-superconducting
quark matter~\cite{Alford:2007xm,Lawley:2006ps}.\par

\begin{acknowledgments}

This research was supported in part by DOE contract DE-AC05-06OR23177,
(under which Jefferson Science Associates, LLC, operates Jefferson
Lab) and in part by the Australian Research Council.  JDC would like
to thank Jefferson Lab for their hospitality and and to thank Ping
Wang for helpful discussions.\par

\end{acknowledgments}

\bibliography{JDCrefs}

\end{document}